\documentclass[aps,prb,twocolumn,showpacs]{revtex4}
\usepackage{epsf}
\usepackage[intlimits]{amsmath}
\usepackage{amssymb}
\usepackage{graphicx}

\newcommand{\Tc}{T_c}
\newcommand{\Tcoh}{T_{\rm coh}}

\newcommand{\TK}{T_{\rm K}}
\newcommand{\Vk}{V_{\bf k}}
\newcommand{\za}{z_{1\bf k}}
\newcommand{\zb}{z_{2\bf k}}
\newcommand{\zab}{z_{1,2\bf k}}

\begin{document}

\title{
Heavy-fermion metals with hybridization nodes:\\
Unconventional Fermi liquids and competing phases
}

\author{Heidrun Weber}
\author{Matthias Vojta}
\affiliation{Institut f\"ur Theoretische Physik,
Universit\"at zu K\"oln, Z\"ulpicher Str. 77, 50937 K\"oln, Germany}
\date{January 25, 2008}

\begin{abstract}
Microscopic models for heavy-fermion materials often assume a local, i.e.,
momentum-independent, hybridization between the conduction band and the local-moment $f$
electrons. Motivated by recent experiments, we consider situations where this neglect of
momentum dependence is inappropriate, namely when the hybridization function has
nodes in momentum space.
We explore the thermodynamic and optical properties of the highly anisotropic heavy Fermi liquid,
resulting from Kondo screening in a higher angular-momentum channel.
The dichotomy in momentum space has interesting consequences:
While e.g. the low-temperature specific heat is dominated by heavy quasiparticles,
the electrical conductivity at intermediate temperatures is carried by
unhybridized light electrons.
We then discuss aspects of the competition between Kondo effect and ordering
phenomena induced by inter-moment exchange:
We propose that the strong momentum-space anisotropy plays a vital role in selecting
competing phases.
Explicit results are obtained for the interplay of unconventional hybridization with
unconventional, magnetically mediated, superconductivity,
utilizing variants of large-$N$ mean-field theory.
We make connections to recent experiments on CeCoIn$_5$ and other
heavy-fermion materials.
\end{abstract}
\pacs{75.20.Hr,75.30.Mb,71.10.Li}

\maketitle


\section{Introduction}

Recent years have seen a revival of research in heavy-fermion metals,
due to the wealth of fascinating phenomena which can be found in these
materials.\cite{flouquet,stewart,hvl,coleman,hewson}
These include non-trivial charge and spin order,
unconventional superconductivity,
non-Fermi-liquid behavior,
as well as quantum criticality beyond the Landau-Ginzburg-Wilson paradigm.
The standard microscopic description of heavy-fermion systems
is based on versions of the Anderson or Kondo lattice models,
consisting of conduction ($c$) electrons and local moments on a regular lattice,
with a spatially local hybridization or Kondo coupling between
$c$-electrons and local moments.
The formation of a heavy Fermi liquid in such a model originates from
the screening of the local moments at low temperatures
through a lattice generalization of the Kondo effect --
this is reasonably well understood, e.g., using slave-particle or
dynamical mean-field approaches.

For some materials, recent experiments\cite{expins,burch} indicate that the assumption
of a local (i.e. momentum-independent) hybridization is insufficient for
a full understanding of the data.
For example, optical conductivity measurements\cite{burch} in
CeMIn$_5$ (M = Co,Ir,Rh) do not show the conventional, frequently observed
hybridization gap,\cite{dordevich,degiorgi,andersepj,hancock,okamura}
but instead have been interpreted in terms of a distribution of gap
values.
Microscopically, a momentum dependence in the hybridization is not surprising,
as the local-moment orbitals are usually of $f$ type and may hybridize
with several conduction-electron orbitals.
While in certain cases this momentum dependence does not lead to significant changes
in observable properties (as compared to a local hybridization), the physics
can be qualitatively different if the hybridization has zeros (i.e. nodes)
in momentum space.
In analogy with unconventional superconductors, having a pair wavefunction
with non-zero internal angular momentum, we may call heavy-fermion materials
``unconventional'' Fermi liquids if the Kondo electron-hole pairs have non-zero
angular momentum.
Such unconventional heavy Fermi liquids, formed below the coherence
temperature in the described setting, will have e.g. quasiparticles with strong
momentum-space anisotropy, to be discussed in more detail in this paper.
(It is worth pointing out that the existence of hybridization nodes does {\em not} imply
that parts of the local moments remain unscreened.)

Importantly, unconventional hybridization will influence the entire complex
phase diagram of heavy-fermion compounds, where the lattice Kondo effect and
various types of long-range order compete for the same electrons.
Clearly, a strong momentum dependence of the hybridzation may favor or
disfavor certain ordering phenomena.
For instance, a dichotomy in momentum space arising from anisotropic
Kondo physics may determine which (unconventional) superconducting phase
is realized at lowest temperatures.

On the theory side, hybridization with higher angular momentum has been discussed
in a few papers only.
Refs.~\onlinecite{ikeda,moreno1} studied the case of a half-filled conduction where a
Kondo semimetal replaces the conventional Kondo insulator -- this physics
is likely relevant to CeNiSn and CeRhSb.\cite{ikeda}
Recently, Ghaemi and Senthil\cite{ghaemi} examined aspects on ``higher
angular-momentum Kondo liquids'', starting from a Kondo lattice
model with non-local Kondo coupling.
Some of their results in the Fermi-liquid regime are related to ours below,
and we shall comment on similarities and differences.
Let us note that differentiation of electronic properties in momentum
space is a common theme in correlated electron systems:
For instance, in the copper-oxide high-temperature superconductors,
quasiparticles properties are known to vary strongly along the Fermi
surface.\cite{fermiarc}

The purpose of this paper is a detailed investigation of heavy-fermion metals
featuring hybridization functions with momentum-space nodes.
In the first part, we study important Fermi-liquid properties including
Fermi surface, effective mass, specific heat, and optical conductivity.
We also discuss the temperature-dependent electrical resistivity.
The focus of the second part is on ordering phenomena competing with
Kondo screening:
Here we concentrate on magnetically mediated superconductivity.
Assuming a direct exchange interaction between local moments, we discuss
mean-field phase diagrams focussing on the interplay of hybridization symmetry
and pairing symmetry.
Most of the concrete calculations are done using the slave-boson mean-field
approximation on two-dimensional lattices,
but most of our ideas apply more generally, including to situations
where the hybridization does not have nodes, but otherwise varies strongly
in momentum space.

The remainder of the paper is organized as follows.
In Sec.~\ref{sec:model} we introduce the microscopic model
to be employed; its mean-field treatment is subject of Sec.~\ref{sec:MF}.
We shall point out the relation between Kondo and Anderson lattice
models with non-local coupling between $c$ and $f$-electrons.
Sec.~\ref{sec:FL} is devoted to properties in the Fermi-liquid
regime, comparing local with non-local (unconventional) hybridization.
In particular, the optical conductivity will be calculated and
discussed in relation to experiments on the CeMIn$_5$ compounds.\cite{burch}
In Sec.~\ref{sec:beyond} we shall touch upon the low-temperature physics
beyond the slave-boson approximation.
Sec.~\ref{sec:rho} discusses qualitative properties of the electrical
resistivity, in particular the perturbatively accessible temperature
regime above the single-impurity Kondo temperature $\TK$.
The competition of Kondo screening and superconducting pairing
is subject of Sec.~\ref{sec:SC}. Two regimes will be distinguished,
depending on whether the transition temperature $\Tc$ is comparable to or
smaller than Fermi-liquid coherence temperature $\Tcoh$.
An outlook will conclude the paper.


\section{Model}
\label{sec:model}

In this paper, we shall restrict our considerations to two-band
models of heavy-electron materials, with simple tight-binding hopping
of electrons.
We shall generalize the Anderson and Kondo lattice models to the
case of a non-local coupling between the conduction ($c$) and local ($f$)
electrons, and discuss the relation between two models.

The Hamiltonian of the Anderson lattice model is given by
\begin{eqnarray}
\label{eq:ALM}
\mathcal{H}_{\text{ALM}}&=&\sum_{{\bf k}\sigma}  (\epsilon_{\bf k}-\mu) c_{{\bf k}
\sigma}^{\dagger}c_{{\bf k} \sigma}^{\phantom{\dagger}} +\sum_{{\bf k} \sigma}
(\epsilon_f-\mu) f_{{\bf k} \sigma}^{\dagger} f_{{\bf k} \sigma} \\
&+&\sum_{{\bf k}
\sigma} V_{{\bf k}}  (f_{{\bf k} \sigma}^{\dagger} c_{{\bf k}
\sigma}^{\phantom{\dagger}}+c_{{\bf k} \sigma}^{\dagger} f_{{\bf k}
\sigma}^{\phantom{\dagger}}) +U \sum_i n_{f,i\uparrow} n_{f,i\downarrow}
\nonumber
\end{eqnarray}
where $c_{{\bf k} \sigma}^{\dagger}$ ($f_{{\bf k} \sigma}^{\dagger}$) creates a
conduction ($f$) electron with momentum ${\bf k}$, spin $\sigma$ and
energy $\epsilon_{\bf k}$ ($\epsilon_f$). $f$- and $c$-electrons are hybridized via the
momentum-dependent hybridization $V_{{\bf k}}$.
Finally, $U$ is the on-site Coulomb repulsion of $f$-electrons.
The chemical potential $\mu$ influences the band fillings
$n_c$ and $n_f$ of the $c$ and $f$-electrons, respectively.
(Note that $\mathcal{H}_{\text{ALM}}$ has the full translational invariance of the
underlying lattice.)

While a full microscopic treatment of the $f$ electron lattice would require
to consider the low-energy Kramers doublet state,
e.g., of one $f$ electron in a $f^1$ configuration of Ce
in the presence of spin-orbit and crystal-field interactions,
we shall proceed with the simplified model \eqref{eq:ALM}.
Hence, we shall take the momentum dependence of $\Vk$ as a phenomenological input,
noting that it is dictated by the lattice structure and the overlap of $c$ and $f$
orbitals.
(In principle, the momentum dependence of $\Vk$ can be further renormalized by
interaction effects, see Sec.~\ref{sec:beyond} below.)
The directional dependence of $\Vk$ as function of ${\bf k}/k$ may be expanded
into spherical harmonics.
Conventionally, one neglects all non-zero angular-momentum components and
assumes a local hybridization $\Vk=V$.
This paper is concerned with situations where the zero angular-momentum component
is small or vanishes, and hence the momentum dependence of $\Vk$ can
no longer be ignored (because e.g. $\Vk$ displays nodes in momentum space).
It is convenient to decompose $\Vk=V \beta_{\bf k}$,
where the form factor $\beta_{\bf k}$ is dimensionless and normalized
to, e.g., $\sum_k \beta_{\bf k}^2 = \mathcal{N}$ where $\mathcal{N}$ is the number of
lattice sites.
It is obvious that the thermodynamics as well as most other observables of the system
will depend on $|\Vk|^2$ only, exceptions will be noted in the course of the paper.
The $f$ level is assumed to be non-dispersive; this approximation
is relaxed in Sec.~\ref{sec:SC}.

In the Kondo limit, i.e., $V\to\infty$, $U\to\infty$, $\epsilon_f\to-\infty$ with
$V^2/\epsilon_f$ finite, charge fluctuations are frozen.
A Schrieffer-Wolff transformation,\cite{hewson} which projects out empty and doubly occupied states
of the $f$ levels, leads to a Kondo lattice model:
\begin{equation}
\label{eq:KLM}
\begin{split}
\mathcal{H}_{\text{KLM}}=&\sum_{{\bf k}\sigma} \bar\epsilon_{\bf k} c_{{\bf k}
\sigma}^{\dagger}c_{{\bf k} \sigma}^{\phantom{\dagger}} +
\underbrace{\sum_{{\bf k} {\bf
k'}i} 2 J_{{\bf k}{\bf k'}} e^{-i ({\bf {\bf k'}}-{\bf k}){\bf R}_i} {\bf S}_i\cdot {\bf
s}_{{\bf k},{\bf k'}}}_{\mathcal{H}_J}
\end{split}
\end{equation}
where ${\bf s}_{{\bf k} {\bf k'}}= \sum_{\sigma \sigma'}c_{{\bf k}
\sigma}^{\dagger} {\bf \sigma}_{\sigma \sigma'} c_{{\bf k'}\sigma'} / 2$,
and $\bar\epsilon_{\bf k}=\epsilon_{\bf k}-\mu$.
In the Kondo limit, $n_f$ is fixed to unity,
and ${\bf S}_i$ is the local moment at site $i$ formed out of the $f$-electrons.
(An additional potential scattering term arising from the Schrieffer-Wolff transformation
will be neglected.)
To leading order, the Kondo coupling is
\begin{equation}
J_{{\bf k}{\bf k'}} = 2 V_{{\bf k}} V_{{\bf k'}} \left(
\frac{1}{U+\epsilon_f}+\frac{1}{-\epsilon_f}\right)
= J_0 \beta_{\bf k} \beta_{\bf k'}
 \label{eq:JK}
\end{equation}
where $J_0=V^2 (\frac{1}{U+\epsilon_f}+\frac{1}{-\epsilon_f})$.
In real space, the Kondo interaction $\mathcal{H}_J$ takes the form
\begin{equation}
\mathcal{H}_J=2 J_0\sum_{imn} \beta_{n-i} \beta_{m-i} {\bf S}_i \cdot {\bf s}_{mn}
\label{kreal}
\end{equation}
where $\beta_{n-i}$ denotes the Fourier transform of $\beta_{\bf k}$ depending on the
distance $|{\bf R}_n-{\bf R}_i|$,
and ${\bf s}_{mn}= \sum_{\sigma \sigma'}
c_{m\sigma}^{\dagger} {\bf \sigma}_{\sigma \sigma'} c_{n\sigma'} / 2$
is a {\em non-local} conduction-electron spin density.

Importantly, for each impurity ${\bf S}_i$,
Eq.~\eqref{kreal} describes a {\em single-channel} Kondo model,
with the exchange ``symmetry'' determined by the hybridization symmetry of the
underlying Anderson model.
In contrast, Ghaemi and Senthil\cite{ghaemi} start out from a Kondo
lattice model where each local moment is exchange-coupled to neighboring
conduction-electron sites as follows:
\begin{equation}
\mathcal{H}'_J=2 \sum_{im} J_{im} {\bf S}_i \cdot {\bf s}_{mm} .
\end{equation}
This is a {\em multi-channel} Kondo model, where the different
screening channels correspond to different linear combinations of the conduction
electrons at the surrounding sites $m$.
In such a model, the screening channels are of different strength,
and the strongest will dominate the low-temperature physics.
The authors of Ref.~\onlinecite{ghaemi} argue that, depending on microscopics (i.e. lattice
and band structure properties), a higher-angular-momentum channel (e.g. $d$-wave) can dominate
over the conventional symmetric ($s$-wave) channel.
The general relation between Anderson and Kondo models with
non-local coupling has been discussed e.g. in Ref.~\onlinecite{ColemanMC}.
There it was argued that, in an Anderson model,
the coupling to a {\em correlated} conduction band opens new screening
channels, and the effective model will be a multi-channel Kondo
model, because the charge fluctuations in the band
(accompanying the non-local hopping) are suppressed by conduction-electron
correlations.

For our purpose, we note that the two Kondo lattice models
with $\mathcal{H}_J$ and $\mathcal{H}'_J$
(with one screening channel dominating) become equivalent in the
slave-boson mean-field (saddle-point) analysis employed below,
because the different screening channels correspond to different
saddle points.
Therefore, our mean-field results derived for the single-channel models
(\ref{eq:ALM},\ref{eq:KLM}) can be directly compared with the ones for the
multi-channel Kondo model in Ref.~\onlinecite{ghaemi}.


\section{Mean-field approximation}
\label{sec:MF}

To obtain quantitative results, we shall employ
the standard slave-boson mean-field approximation
for the Kondo and Anderson lattice models (the latter with infinite $U$).
In the following, we briefly summarize the corresponding
formalism.\cite{hewson,readnewns,burdin}
Note that in all cases we restrict our attention to states with spatial
translational invariance.

We start with the Anderson model.
For infinite on-site repulsion, the three states of each $f$ orbital
can be represented by auxiliary fermions $\bar f_{i\sigma}$ and spinless bosons $r_i$,
such that the physical $f$ electrons $f_{i\sigma} = r_i^\dagger \bar f_{i\sigma}$,
together with the constraint
\begin{equation}
\label{eq:constraint} \sum_{\sigma}\bar f_{i \sigma}^{\dagger}\bar f_{i
\sigma}^{\phantom{\dagger}}+ r_{i}^{\dagger}r_{i}^{\phantom{\dagger}}=1 \,.
\end{equation}
At the saddle point, the slave bosons condense, $\langle r_{i}\rangle=r$, which
implies a rigid hybridization between the $c$ and the $\bar f$ bands.
The mean-field Hamiltonian of the Anderson lattice model reads
\begin{equation}
\label{eq:H_MFALM}
\begin{split}
\mathcal{H}&_{\text{ALM,MF}}=\sum_{{\bf k}\sigma}  \epsilon_{\bf k} c_{k
\sigma}^{\dagger}c_{{\bf k} \sigma} +\sum_{{\bf k} \sigma} \epsilon_f
\bar f_{{\bf k} \sigma}^{\dagger} \bar f_{{\bf k} \sigma}\\
&+\sum_{{\bf k} \sigma} V_{\bf k} r (\bar f_{{\bf k} \sigma}^{\dagger} c_{{\bf k}\sigma}
+c_{{\bf k} \sigma}^{\dagger} \bar f_{{\bf k}\sigma})\\
&-\lambda \left(\sum_{{\bf k} \sigma}  \bar f_{{\bf k}
\sigma}^{\dagger} \bar f_{{\bf k} \sigma} + \mathcal{N}(r^2-1)\right)
-\mu \left(\sum_{{\bf k}\sigma}  c_{{\bf k} \sigma}^{\dagger}c_{{\bf k}
\sigma}-\mathcal{N} n_c\right)
\end{split}
\end{equation}
where $\lambda$ is the Lagrange multiplier implementing the constraint (\ref{eq:constraint})
at the mean-field level,
and the effect of the chemical potential $\mu$ on the $\bar f$ electrons has been absorbed in $\lambda$.
The three mean-field parameters $r$, $\lambda$, and $\mu$ are obtained
from minimizing the free energy, leading to the self-consistency equations:
\begin{subequations}
\label{eq:MF_ALM}
\begin{eqnarray}
\sum_{{\bf k}\sigma}V_{{\bf k}} \langle \bar f_{{\bf k} \sigma}^{\dagger} c_{{\bf
k} \sigma}^{\phantom{\dagger}}+ h.c. \rangle &=&2 \mathcal {N} \lambda r , \\
\sum_{{\bf k} \sigma}  \langle \bar f_{{\bf k} \sigma}^{\dagger} \bar f_{{\bf k}
\sigma}^{\phantom{\dagger}}\rangle &=& \mathcal{N} (1-r^2) , \\
\sum_{{\bf k}\sigma} \langle c_{{\bf k} \sigma}^{\dagger} c_{{\bf k}
\sigma}^{\phantom{\dagger}} \rangle &=& \mathcal{N} n_c .
\end{eqnarray}
\end{subequations}
The expectation values can be easily expressed in terms of the Green's function
of the diagonalized mean-field Hamiltonian, for details
see App.~\ref{sec:GF}.

For the mean-field analysis of the Kondo lattice model one represents
the local moments ${\bf S}_i$ by auxiliary fermions $\widetilde f_{i \sigma}$,
${\bf S}_i=\frac{1}{2}\sum_{\sigma\sigma'} \widetilde f_{i \sigma}^{\dagger} {\bf
\sigma}_{\sigma \sigma'}\widetilde f_{i\sigma'}^{\phantom{\dagger}}$,
with the constraint
\begin{equation}
\sum_{\sigma} \widetilde f_{i \sigma}^{\dagger}\widetilde f_{i
\sigma}^{\phantom{\dagger}} = 1\,.
\end{equation}
The Kondo interaction takes the form
\begin{equation}
\mathcal{H}_J=-J_0\sum_{imn\sigma \sigma'} \beta_{n-i} \beta_{m-i}\widetilde f^{\dagger}_{i \sigma}
c^{\phantom{\dagger}}_{n\sigma} c^{\dagger}_{m\sigma'}\widetilde f^{\phantom{\dagger}}_{i\sigma'}
\label{eq:HJ}
\end{equation}
where additional bilinear terms have been dropped, as they can be absorbed
in chemical potentials.
The Kondo interaction term $\mathcal{H_J}$ can be decoupled using
auxiliary fields $b_i$ conjugate to $(-J_0 \sum_{n\sigma} \beta_{n-i}
\widetilde f_{i\sigma}^{\dagger} c_{n\sigma}^{\phantom{\dagger}})$,
i.e., $b_i$ reflects the hybridization between the $\widetilde f$ and $c$ bands
at site $i$.
At the saddle point, the $b_i$ condense, and translational invariance dictates $b_i=b$.
The Kondo lattice mean-field Hamiltonian is
\begin{equation}
\begin{split}
\mathcal{H}&_{\text{KLM,MF}}= \sum_{{\bf k}\sigma} \epsilon_{\bf k} c_{{\bf k}
\sigma}^{\dagger}c_{{\bf k} \sigma}^{\phantom{\dagger}}\\ &+b \sum_{{\bf k} \sigma}
\beta_{\bf k} \left(c_{{\bf k} \sigma}^{\dagger}\widetilde  f_{{\bf k}
\sigma}^{\phantom{\dagger}}+h.c.\right)+\mathcal{N}\frac{b^2}{J_0}\\ &-\lambda_0
\left(\sum_{{\bf k} \sigma}\widetilde  f_{{\bf k} \sigma}^{\dagger}\widetilde f_{{\bf k}
\sigma}^{\phantom{\dagger}}-\mathcal{N}\right) -\mu \left(\sum_{{\bf k}\sigma}  c_{{\bf
k} \sigma}^{\dagger}c_{{\bf k} \sigma}^{\phantom{\dagger}}-\mathcal{N} n_c\right)
\end{split}
\label{eq:H_KLMMF}
\end{equation}
As before, the three parameters $b$, $\lambda_0$, and $\mu$ are
determined by self-consistency equations which now read:
\begin{subequations}
\label{eq:MF_KLM}
\begin{eqnarray}
\sum_{{\bf k}\sigma} \beta_{\bf k} \langle \widetilde f_{{\bf k} \sigma}^{\dagger} c_{{\bf k}
\sigma}^{\phantom{\dagger}}+ h.c. \rangle &=& -\mathcal{N} \frac{2 b}{J_0} , \\
\sum_{{\bf k} \sigma} \langle \widetilde  f_{{\bf k} \sigma}^{\dagger}\widetilde f_{{\bf k}
\sigma}^{\phantom{\dagger}}\rangle &=& \mathcal{N} , \\
\sum_{{\bf k}\sigma} \langle c_{{\bf k} \sigma}^{\dagger} c_{{\bf k}
\sigma}^{\phantom{\dagger}} \rangle &=& \mathcal{N} n_c .
\end{eqnarray}
\end{subequations}
These mean-field equations are equivalent to the ones of the Anderson lattice model, Eq.~(\ref{eq:MF_ALM}),
if the Kondo limit is taken there, for details see App.~\ref{sec:equiv}.

We note that mean-field Hamiltonians (\ref{eq:H_MFALM},\ref{eq:H_KLMMF}) represent
the $N=\infty$ saddle-point solutions of certain SU($N$) Anderson and Kondo lattice models.
In this mean-field picture, both the Anderson and Kondo lattice models are
mapped onto two-band systems of non-interacting fermions
with a self-consistently determined renormalized hybridization between the bands.
At high temperature, the slave-boson condensation amplitude vanishes
(leading to two decoupled bands), whereas the condensation amplitude is
finite below the single-impurity Kondo temperature $\TK$.
In the Kondo lattice model, $\TK$ is given by:
\begin{equation}
\frac{2}{J_0}= \frac{1}{\mathcal{N}}\sum_{\bf k}
\frac{\beta^2_{\bf k}}{\bar \epsilon_{\bf k}} \tanh \frac{\bar \epsilon_{\bf k}}{2\TK}
\end{equation}
with $\bar \epsilon_{\bf k}=\epsilon_{\bf k}-\mu$.
The neglect of fluctuations in the mean-field approach
causes an artificial finite-temperature transition
at $\TK$ which can in principle be
cured by including the coupling to a U(1) gauge field, see Sec.~\ref{sec:beyond}.

The effective two-band picture is appropriate to
describe the low-temperature Fermi-liquid regime, i.e., the quasiparticle physics
for temperatures below the Fermi-liquid coherence temperature $\Tcoh$.
An approximation for $\Tcoh$ can be extracted from the $T\!=\!0$ slave-boson solution:
For the Kondo lattice model one obtains $\Tcoh = b^2/D$ where
$D$ is the bandwidth.\cite{burdin}
(Also, $\lambda_0 \propto \Tcoh$.)
The Fermi surface resulting from the slave-boson approximation
fulfills Luttinger's theorem:
The momentum-space volume enclosed by the Fermi surface is proportional
to the total number of electrons $n_{\rm tot} = n_c+n_f$ where
$n_f=1$ in the Kondo limit.
At small non-zero temperature $T$, the mean-field parameters
acquire quadratic $T$ corrections characteristic of a Fermi liquid,
e.g., $[b(T)-b_0]/b_0 \propto -T^2/T_K^2$ where $b_0=b(T\!=\!0)$.

The simple two-band picture of the slave-boson approach has been confirmed, e.g.,
using the dynamical mean-field theory\cite{dmft-rmp} (DMFT) approach to
the Anderson lattice model, which fully includes local correlations and inelastic
processes:
The DMFT results\cite{grenzebach} nicely show the formation of a coherent heavy band
crossing the Fermi level at low temperatures, rather well separated from the second
band.


\section{Low-temperature properties of the Fermi-liquid state}
\label{sec:FL}

This section will discuss the properties of the heavy-Fermi liquid
state in a heavy-fermion system with unconventional hybridization.
Effects of inter-moment exchange and possible ordered phases
will be ignored, we will come back to these issues in Sec.~\ref{sec:SC}.
Quantitative results will be obtained by solving the slave-boson mean-field
equations for different model parameters,
but the qualitative aspects will be of general validity unless otherwise noted.

We restrict ourselves to two-dimensional systems on a square lattice.
For the $c$-electrons a tight-binding dispersion will be assumed:
\begin{equation}
\epsilon_{\bf k}=-2 t\left(\cos k_x+\cos k_y\right).
\label{2dtight}
\end{equation}
Results will be shown for hybridization functions of the form
$\Vk = V_0 \beta_{\bf k}$ with
\begin{equation}
\beta_{\bf k} = \left\{
\begin{array}{ll}
1 & \text{$s$-wave}\\ \cos k_x+\cos k_y & \text{extended $s$-wave}\\ \cos k_x-\cos k_y &
\text{$d_{x^2-y^2}$-wave}\\
\end{array}
\right.
\label{eq:symm}
\end{equation}
The $d_{x^2-y^2}$-wave and the extended $s$-wave case correspond to different linear combinations
of hybridization between an $f$-site and its nearest-neighbor $c$-sites
(also discussed by Ghaemi and Senthil\cite{ghaemi}),
while the $s$-wave case with local hybridization is shown for comparison.

In addition, we will also consider a lattice appropriate to model the CeIn planes of
the CeMIn$_5$ materials (M=Ir, Rh or Co). Those crystallize in a HoCoGa$_5$-type
tetragonal structure,\cite{structureCeMIn5,lda115} where the Ce and the in-plane In ions
are located on two interpenetrating square lattices.
Thus we assume a $c$ electron dispersion as in Eq.~\eqref{2dtight},
and hybridization functions of the form
\begin{equation}
\beta_{\bf k} = \left\{
\begin{array}{ll}
2 \cos \frac{k_x}{2} \cos \frac{k_y}{2} & \text{extended $S$-wave}\\
2 \sin \frac{k_x}{2} \sin \frac{k_y}{2}& \text{$D_{xy}$-wave}\\
\end{array}
\right..
\label{eq:symm2}
\end{equation}

Hybridization functions $\beta_{\bf k}$ which formally break inversion symmetry,
i.e. have odd angular momentum $l$, are possible as well.
Observables are governed by $V_k^2$ and display inversion symmetry;
hence there is little qualitative difference between even and odd angular momentum
hybridization (apart from the location of the hybridization nodes).
Exceptions are transport anisotropies for $l=1$, briefly discussed in
Sec.~\ref{sec:thermal}.

\subsection{Band structure and Fermi surface}

\begin{figure}[tb]
\resizebox{240pt}{!}{
\includegraphics{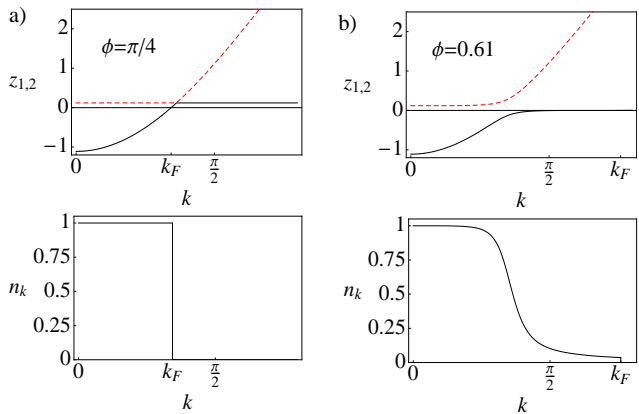}
}
\caption{
(Color online)
Dispersion relation of the two effective bands, $\zab$ \eqref{eq:excit},
and the corresponding momentum distribution function $n_{\bf k}$,
for a $d_{x^2-y^2}$-wave hybridization on a 2d square lattice,
a) along the momentum-space diagonal,
b) along a direction which encloses the angle $0.195\pi$ with the $k_x$ axis.
The parameters of the Kondo lattice model are $J_0/t=2.0$ and $n_c=0.4$.
}
\label{fig:disp}
\end{figure}

\begin{figure*}[tbh]
\resizebox{450pt}{!}{
\includegraphics{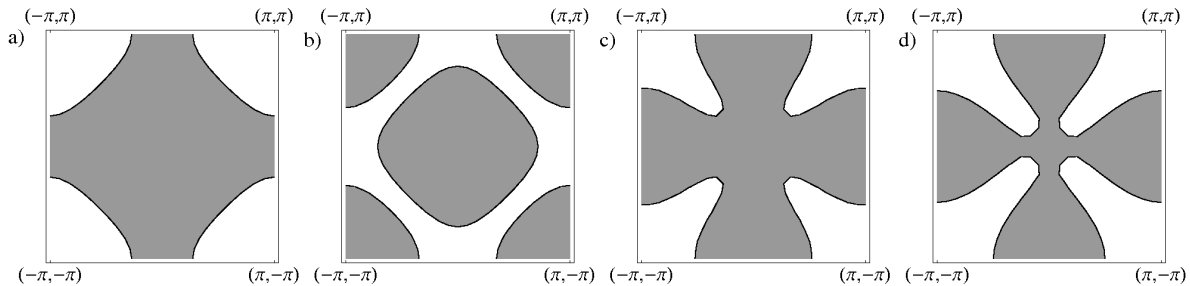}
}
\caption{
Fermi surfaces for
a) $s$-wave,
b) extended $s$-wave, and
c,d) $d_{x^2-y^2}$-wave
hybridization.
The band filling is $n_c=0.3$ in panels a--c, whereas $n_c=0.1$ in panel d;
the Kondo coupling is chosen such that the specific heat coefficient
is identical in all four cases:
a) $J_0/t=1.0$    ($\lambda_0=-0.025$,  $\mu=-2.47$,  $b=0.281$),
b) $J_0/t=0.89$  ($\lambda_0=-0.0067$, $\mu=-2.3$,   $b=0.281$),
c) $J_0/t=0.97$  ($\lambda_0=-0.0061$, $\mu=-2.46$,  $b=0.173$),
d) $J_0/t=1.957$ ($\lambda_0=-0.0385$, $\mu=-3.698$, $b=0.466$).
} \label{fig:FF}
\end{figure*}

To simplify the discussion, we shall work in the Kondo limit.
The eigenvalues of the mean-field Hamiltonian (\ref{eq:H_KLMMF}),
representing the effective bands of the heavy Fermi liquid,
are given by
\begin{equation}
\zab = \frac{1}{2} \left( -\lambda_0+ {\bar \epsilon_{\bf k}}
\pm \sqrt{(\lambda_0+ {\bar \epsilon_{\bf k}})^2+4 b^2 \beta_{\bf k}^2}\right).
\label{eq:excit}
\end{equation}
This band structure is illustrated in Fig.~\ref{fig:disp}.
Along certain ``nodal'' lines in momentum space the hybridization vanishes,
and the two bare bands of the $c$ and $\widetilde f$ particles cross (Fig.~\ref{fig:disp}a);
for $d_{x^2-y^2}$ symmetry this applies to $k_x=\pm k_y$.
Otherwise, the hybridization causes a band repulsion (Fig.~\ref{fig:disp}b),
which is maximum in the ``antinodal'' direction.
As the bare $\widetilde f$ band is non-dispersive, the two bands $\zab$
do not overlap, and consequently only one band crosses the Fermi level.
For less than half filling, $n_c < 1$, the Fermi surface is thus
determined by $\zb=0$.

Fig. \ref{fig:disp} also shows the momentum distribution function of the
$c$-electrons,
$n_{\bf k}=\langle c_{\bf k}^{\dagger} c_{\bf k}^{\phantom{\dagger}}\rangle$.
It shows a jump at the Fermi wave vector $k_F$, the jump height given by
the quasiparticle weight $Z$, see below. As typical for heavy-fermion systems,
$n_{\bf k}$ also shows a rounded step at the ``small'' Fermi surface
of the original $c$ electrons, i.e., the the Fermi surface
in the absence of a hybridization (or at temperatures $T\gg\Tcoh$).

Sample results from a full numerical solution of the mean-field equations (\ref{eq:MF_KLM})
are shown in Figs. \ref{fig:FF}, \ref{fig:FFrot} and in
Figs.~\ref{fig:grad}, \ref{fig:Z} below.
Fig. \ref{fig:FF} displays 2d Fermi surfaces for the hybridization functions
of Eq.~\eqref{eq:symm}, with parameters chosen such that the specific heat
coefficient is identical in the four cases, see below.
For $d_{x^2-y^2}$-wave hybridization, Figs.~\ref{fig:FF}c,d,
one clearly observes a ``small'' Fermi momentum $k_F$ along the nodal directions,
while $k_F$ is ``large'' along the antinodal direction.
Note that the function $k_F(\phi)$,
where $\phi$ is the angle parameterizing the $k$ space direction,
can be multivalued due to the momentum dependence of $\Vk$:
Analyzing the equation $\zb=0$ one finds that this generically
happens for small band filling $n_c$ (Fig.~\ref{fig:FF}d).
In the case of extended $s$-wave hybridization, Fig.~\ref{fig:FF}b,
two Fermi sheets emerge, as the lower band crosses the Fermi level twice.

For the hybridizations of Eq.~\eqref{eq:symm2}, arising from two interpenetrating
square lattices of $c$ and $f$ electrons, sample Fermi surfaces are shown in
Fig.~\ref{fig:FFrot}.
In the $d$-wave case, Fig.~\ref{fig:FFrot}b, the nodal lines are simply rotated
by 45 degrees w.r.t. to Fig.~\ref{fig:FF}c,d.

\begin{figure}[b!]
\resizebox{220pt}{!}{
\includegraphics{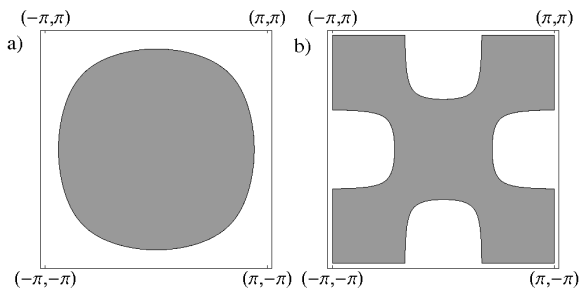}
}
\caption{
Fermi surfaces using the hybridization functions \eqref{eq:symm2} for
interpenetrating $c$ and $f$ square lattices.
a) Extended $S$-wave with $J_0/t=0.46$ ($\lambda_0/t=-0.011$, $\mu/t=-2.38$, $b/t=0.187$),
b) $D_{xy}$-wave with $J_0/t=1.975$) ($\lambda_0/t=-0.001$, $\mu/t=-2.34$, $b/t=0.138$).
In both cases, $n_c=0.3$, and the parameters are chosen such that
the $\gamma$ coefficient is the same as for the data in Fig.~\ref{fig:FF}.
}
\label{fig:FFrot}
\end{figure}

\begin{figure*}[bt]
\resizebox{500pt}{!}{
\includegraphics{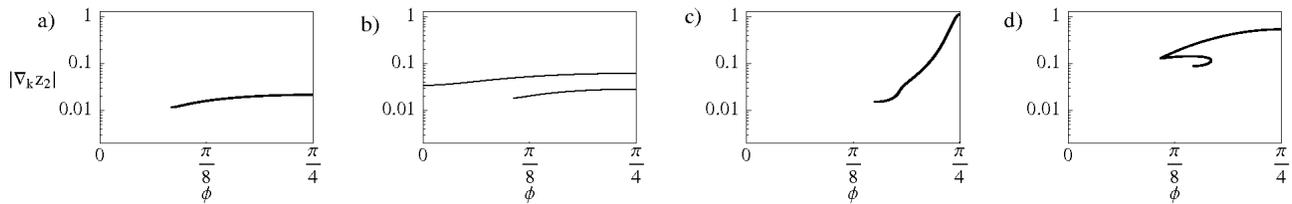}
}
\caption{
The quasiparticle velocity at the Fermi level, $|\nabla_{{\bf k}} \zb|$,
in a logarithmic plot vs. momentum-space angle $\phi$ for
a) $s$-wave,
b) extended $s$-wave, and
c,d) $d_{x^2-y^2}$-wave hybridization.
Parameters are as in Fig.~\ref{fig:FF}a-d.
In all four cases, the ``total'' effective mass (as derived from the specific heat)
is around 125 times the bare electron mass.
}
\label{fig:grad}
\end{figure*}

\begin{figure*}[tb]
\resizebox{500pt}{!}{
\includegraphics{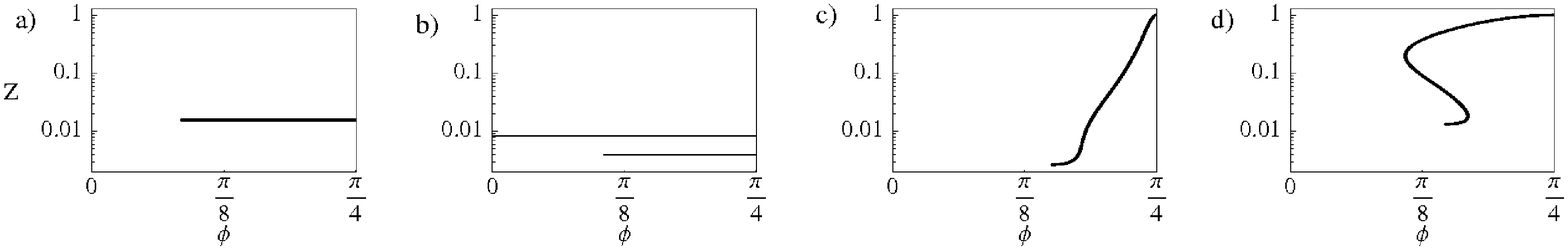}
}
\caption{
As in Fig.~\ref{fig:grad}, but now showing the
quasiparticle weight $Z$ in a logarithmic plot vs. angle $\phi$.
}
\label{fig:Z}
\end{figure*}

\subsection{Thermodynamic properties}
\label{sec:thermprop}

The leading low-temperature thermodynamics can be directly obtained
from the effective two-band description of the slave-boson approximation.\cite{heatfoot}
For non-interacting fermions, the Sommerfeld coefficient, $\gamma=C_V/T$,
of the specific heat is related to the density of states (per spin) at the Fermi level,
$N_0=N(\omega\!=\!0)$, through $\gamma=\frac{2\pi^2}{3} N_0$.
In standard isotropic Fermi-liquid theory, the effective mass $m^\ast$ is defined through
the slope of the dispersion at the Fermi level, $v_F = k_F/m^\ast$,
where $v_F$ and $k_F$ are Fermi velocity and Fermi momentum, respectively.
The density of states is $N_0 = m^\ast k_F^{d-2} / \mathcal{C}_d$
in $d$ dimensions where $\mathcal{C}_2 = 2\pi$ and $\mathcal{C}_3 = 2\pi^2$.

In anisotropic systems, a suitable definition for a (direction-dependent!)
effective mass is:
\begin{equation}
\frac{1}{m^\ast(\bf k)} =
\frac 1 k \left| \frac{\partial\epsilon_{\bf k}}{\partial \bf k}\right|_{\rm FS}
\end{equation}
where $\epsilon_{\bf k}$ is the quasiparticle energy of the band crossing the Fermi
surface (FS), and $k=|{\bf k}|$.
Then, the density of states is given by
\begin{equation}
N_0 = \int_{\rm FS} \frac{d^{d-1}k}{(2\pi)^d} \, \frac{m^\ast(\bf k)}{k} \,.
\end{equation}

Note that Ref.~\onlinecite{ghaemi} defined a quantity $m^\ast$ via
the second (instead of the first) derivative of quasiparticle energies,
which in general plays only a subleading role in thermodynamics.

In our two-band system, $\epsilon_{\bf k} = \zb$ for $n_c<1$.
In $d=2$ some factors of $k$ drop out, such that
\begin{equation}
N_0 = \frac{m^\ast}{2\pi}=
\frac{1}{4 \pi^2} \int_{\text{FS}} \frac{d A_{\bf k}}{|\nabla_{{\bf k}} \zb|}
\end{equation}
where $m^\ast$ is the effective mass as extracted from specific-heat measurements
(i.e. the effective mass of an isotropic Fermi liquid with the same $\gamma$),
and $d A_{\bf k}$ is the Fermi surface element.
Under the condition that the
Fermi surface can be parameterized in the way $k=k_F(\phi)$ this integral can be
rewritten as
\begin{equation}
\label{eq:meff}
N_0 = \frac{m^\ast}{2\pi}=\frac{1}{4 \pi^2} \int \text{d}
\phi\frac{\sqrt{(k_F(\phi))^2+(k_F'(\phi))^2}}{|\nabla_{{\bf k}} \zb|_{k=k_F(\phi)}}.
\end{equation}

Sample results for the quasiparticle velocity, $|\nabla_{{\bf k}} \zb|_{\rm FS}$,
as function of the Fermi-surface angle (i.e. the direction) are shown in
Fig.~\ref{fig:grad} for the various hybridization cases introduced in
Eq.~\eqref{eq:symm}.
The microscopic parameters are chosen such that all four cases lead to the same
value of the density of states $N_0$ and thus the same specific-heat coefficient.
The corresponding total effective mass is around 125 times the bare $c$-electron
mass.
For $d_{x^2-y^2}$-wave hybridization the velocity and
therefore also the inverse effective mass has a maximum at the nodal line ($\phi=\pi/4$) --
here $m^\ast(\bf k)$ corresponds to approximately the bare $c$-electron mass.
Away from the nodal line the velocity rapidly decreases.
In contrast, for both $s$-wave-like hybridizations, the velocity is
approximately constant (and small) along the Fermi surface.

The electronic quasiparticle weight, $Z(\bf k)$, can be easily extracted as well.
$Z$ measures the overlap between the physical $c$ electron and the low-energy
quasiparticle at the Fermi surface.
In the mean-field approach of two hybridized bands, $Z$ is given by:\cite{ghaemi}
\begin{equation}
Z({\bf k}) =
\frac{(\za-\bar \epsilon_{\bf k})^2}{(\za-\bar \epsilon_{\bf k})^2+b^2 \beta_{\bf k}^2}\,.
\end{equation}
Results for $Z$ are displayed in Fig.~\ref{fig:Z}:
It is unity along the nodal lines of the hybridization, but becomes very small
away from it -- the latter is the typical heavy-fermion situation.
For the $s$-wave-like hybridizations, $Z$ turns out to be independent of the
momentum direction; in the $s$-wave case this follows from $\Vk={\rm const}$,
whereas in the extended $s$-wave case this follows from the coincidence
of the momentum dependence of the hybridization $\Vk$ and the $c$ electron dispersion
$\epsilon_{\bf k}$.

At this point, let us comment on a few important issues.
First, even in the presence of hybridization nodes, all local moments of the Anderson or Kondo
lattice are {\em fully screened} in the low-temperature limit.
This is obvious from the slave-boson solution which clearly describes a Fermi liquid,
but also beyond slave bosons we see no reason for a (partial) breakdown of
Kondo screening:
For instance, in DMFT complete screening will occur once the effective bath density
of states at the Fermi level is finite (which is the case here).
Physically, the local moments are entities in real space, whereas the hybridization
nodes are defined in momentum space.
Second, as the nodes cover only a set of momenta of zero measure,
hybridization nodes do not easily lead to so-called two-fluid behavior
(i.e. a heavy-Fermi liquid coexisting with local moments),
which has been advocated on phenomenological grounds.\cite{twofluid}
We note that these statements also hold if both quasiparticle bands
($c$-like and $f$-like) cross the Fermi level.
Although one may speculate about the existence of gapless spinons
at the $f$ Fermi points where the hybridization vanishes,\cite{ghaemi} these
would again cover only a set of momenta of zero measure.

\subsection{Influence of a magnetic field}

Let us briefly discuss the effects of a weak external field applied to the heavy
Fermi liquid.
In general, a Zeeman field will cause a spin splitting of the Fermi surface,
with spin- and field-dependent effective masses [or densities of states
$N_\sigma(B)$].\cite{beach}
The qualitative field dependence, $N_\sigma(B) = N_0(1 + \sigma B/B_0)$ with
$B_0$ proportional to the Kondo temperature, is not changed by a momentum dependence
of $\Vk$.
(The field dependence of the mean-field parameters leads to a subleading
correction $\propto B^2$ to $N_\sigma(B)$.\cite{beach})
The anisotropy of $\Vk$ of course causes an anisotropy of the $k$-space distance
between the spin-split Fermi sheets, as the Fermi velocity is highly anisotropic.

For magneto-oscillation measurements, the cyclotron mass is an important quantity,
given by $2\pi m_c= \partial A(E)/\partial E$,
where $A(E)$ is the area enclosed by the quasiparticle iso-energy curve
in a momentum-space plane perpendicular to the applied orbital field.
Thus, in dimensions $d>2$ the cyclotron mass $m_c$ depends on the field direction.
However, in 2d this dependence is absent, and $m_c$ is identical to the
(averaged) quasiparticle mass $m^\ast$ extracted from the density of states or specific heat,
independent of momentum-space anisotropies.

\subsection{Optical conductivity}
\label{sec:opcon}

The optical response of heavy-fermion metals has been studied extensively.\cite{degiorgi}
Experiments probing the optical conductivity $\sigma(\omega)$ usually show a Drude peak
well separated from mid-infrared excitations.
These features have been interpreted in the two-band picture advocated above:
While intra-band particle-hole excitations produce conventional metallic Drude-like response,
inter-band excitations lead to finite weight at elevated energies.
In a picture of free fermions, the threshold energy of these optical inter-band excitations
measures the minimum gap between occupied and unoccupied states in the lower and upper band,
respectively.
For momentum-independent hybridization between $c$ and $f$ bands,
this optical gap $\Delta_{\rm opt}$
is simply given by twice the value of the renormalized hybridization.
As explained above, the hybridization is expected to scale as the square root of the coherence
temperature, hence $\Delta_{\rm opt} \sim \sqrt {\Tcoh D}$
(where $D$ is the conduction-electron bandwidth).
Clearly, the simple two-band picture falls short of capturing inelastic processes
at non-zero energies, which will inevitably smear out the gap even at $T=0$.
Nevertheless, a pseudogap-like feature has been shown to survive in $\sigma(\omega)$
when fully accounting for dynamic local correlation effects in the framework
of DMFT for the standard Anderson lattice model at large $U$.\cite{grenzebach,fbapriv}
In the results of these calculations, the magnitude of the pseudogap has the same
scaling as above, $\Delta_{\rm opt} \sim \sqrt {\Tcoh}$.
Remarkably, this relation between optical gap and coherence temperature has
been found to be nicely obeyed by a number of heavy-fermion
metals.\cite{dordevich,degiorgi,andersepj,hancock,okamura}
However, optical conductivity studies in CeMIn$_5$ (M=Ir,Rh or Co) show little signatures
of a well-defined hybridization gap.\cite{burch}
As we show below, such a behavior is in principle consistent with a strongly
momentum-dependent hybridization in the underlying Anderson lattice model.

The finite-frequency part of the optical conductivity $\sigma(\omega)$ can be
expressed through the retarded current-current correlation function as
\begin{equation}
\sigma(\omega)=\frac{i}{\omega} \Pi(\omega+i\eta)
\label{eq:sigma}
\end{equation}
with
\begin{equation}
\Pi(i \omega)=-\int_0^{\beta} \text{d}\tau e^{i \omega \tau} \langle T_{\tau} {\bf j}^{\dagger}(\tau) {\bf j}(0) \rangle
\end{equation}
The current operator $\bf j$ has to calculated as the time derivative of the polarization operator $\bf P$
\begin{equation}
{\bf j}=i \left[ \mathcal{H},\bf{P} \right]
\label{eq:current}
\end{equation}
where the definition of $\bf P$ includes all charged particles $a_i$ (with charge $q_i$):
\begin{equation}
{\bf P}=\sum_{i \sigma} q_i {\bf R}_i a_i^{\dagger} a_i^{\phantom{\dagger}} \,.
\end{equation}
As usual, approximations to the propagators and to the current vertex
in calculating $\Pi(\omega)$ have to be mutually consistent in order to respect charge
conservation (expressed by the corresponding Ward identity).

\begin{figure}[tb]
\resizebox{230pt}{!}{
\includegraphics{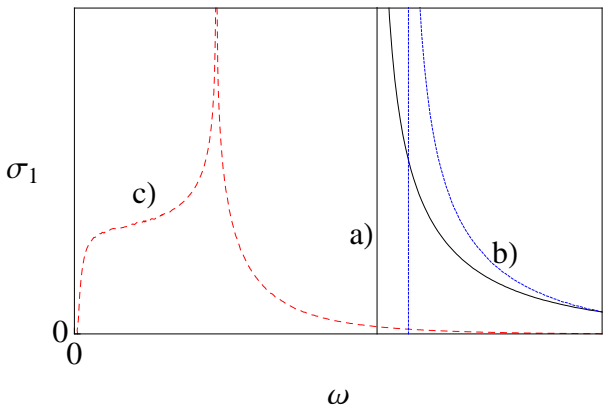}
}
\caption{
(Color online)
Real part of the optical conductivity $\sigma_1(\omega)$ for
a) $s$-wave (solid),
b) extended $s$-wave (dotted), and
c) $d_{x^2-y^2}$-wave hybridization (dashed).
The parameters are as in Fig.~\ref{fig:FF}a-c,
for a discussion see text.
}
\label{fig:oplfk}
\end{figure}

At this point, the electrodynamics of the heavy Fermi liquid requires a
thorough discussion. Physically, the $f$ electrons contribute to the
Fermi surface and carry charge.
While this is plausible in the Anderson model picture,
where the charge is naturally carried by the $\bar f$ auxiliary particles,
the Kondo case is more subtle:
The $\widetilde f$ particles of the mean-field theory are neutral spinons,
which will carry a physical electric charge only upon inclusion of gauge fluctuations,
see Sec.~\ref{sec:beyond}.
Hence, we shall take the Anderson model viewpoint here.
In the spirit of the mean-field theory, we demand the current correlator to be calculated
as the bare bubble. An expression for the current vertex, which is consistent with the
mean-field propagators, is obtained from:
\begin{equation}
{\bf j}_{\text{MF}}=i \left[ \mathcal{H}_{\text{ALM,MF}},\mathbf{P}_{\rm MF} \right]
\label{eq:current2}
\end{equation}
where charge-carrying particles $c_i$ and $\bar f_i$ are contained in ${\bf P}_{\rm MF}$:
\begin{equation}
{\bf P}_{\rm MF} =
\sum_{i \sigma} {\bf R}_i\left( c_{i \sigma}^{\dagger} c_{i \sigma} +\bar f_{i \sigma}^{\dagger} \bar f_{i \sigma} \right)
\end{equation}
Evaluating Eq.~(\ref{eq:current2}) leads to
\begin{equation}
{\bf j}_{\text{MF}}=\sum_{{\bf k} \sigma}\left( (\nabla_{{\bf k}} \epsilon_{\bf{k}}) c_{{\bf k} \sigma}^{\dagger} c_{{\bf k} \sigma}^{\phantom{\dagger}}
+r (\nabla_{\bf k} V_{\bf k}) (c_{{\bf k} \sigma}^{\dagger} \bar f_{{\bf k} \sigma}+h.c.)\right).
\label{eq:jmf}
\end{equation}
Let us pause to emphasize that a current operator derived from $\mathcal{H}_{\text{ALM}}$
{\em before} the mean-field approximation would have a $f$ electron contribution
different from that in ${\bf j}_{\text{MF}}$, but such a current vertex would
be {\em inconsistent} when used together with the bubble of mean-field propagators, i.e.,
vertex corrections would become important.
We point out that ${\bf j}_{\text{MF}}$ has several shortcomings
because ${\bf P}_{\rm MF}$ treats the $\bar f$ as real electrons,
nevertheless, the expression \eqref{eq:jmf} is the {\em only} current operator suitable within
the mean-field treatment of the Anderson model.
We also note that the second term in ${\bf j}_{\text{MF}}$ vanishes in the conventional case
of a constant hybridization $\Vk$, and ambiguities regarding the treatment of the $f$
electrons do not arise.\cite{grenzebach,czycholl}

Using the current operator ${\bf j}_{\text{MF}}$ \eqref{eq:jmf},
the expression for the real part of the optical conductivity $\sigma_1$
(still assuming $n_c<1$) reads
\begin{equation}
\label{eq:sigma1}
\sigma_1(\omega)=\frac{\pi}{\omega} \sum_{\bf k} \frac{n_F(z_{2 {\bf k}})}{( z_{1 {\bf k}}-z_{2 {\bf k}})^2}
A_{\bf k} \delta(z_{1 {\bf k}}-z_{2 {\bf k}}-\omega)
\end{equation}
using the abbreviations $\widetilde \epsilon_f=\epsilon_f-\lambda$ and
\begin{equation}
\begin{split}
A_{\bf k} =&
\left(\nabla_{\bf k} \epsilon_{\bf k} \right)^2 (z_{2 {\bf k}}-\widetilde \epsilon_f) (z_{1 {\bf k}}-\widetilde \epsilon_f)+\\
&+r^2 \left(\nabla_{\bf k} V_{\bf k} \right)^2 \left( (z_{1 {\bf k}}-z_{2 {\bf k}})+4 r^2 V_{\bf k}^2 \right)+\\
&+2 r^2 V_{\bf k} \nabla \epsilon_{\bf k} \nabla V_{\bf k} \left(z_{2 {\bf k}}+z_{1 {\bf k}}-2 \widetilde \epsilon_f \right)
\end{split}
\end{equation}

The result for $\sigma(\omega\!>\!0)$ from a numerical evaluation of Eq.~(\ref{eq:sigma1})
is depicted in Fig.~\ref{fig:oplfk}, for the three hybridization symmetries of
Eq.~\eqref{eq:symm} and parameters as in Fig.~\ref{fig:FF}a-c.
In all situations, a finite gap $\Delta_{\rm opt}$ is visible in $\sigma(\omega)$, which corresponds to the
minimal direct gap between the two bands $\za$ and $\zb$.
In the cases of the $s$-wave and extended $s$-wave hybridization, $\Delta_{\rm opt}$ is
given by $2 b$ and $|\lambda_0-\mu| b/\sqrt{b^2+t^2}$, respectively;
both expressions translate into $\Delta_{\rm opt} \sim \sqrt{\Tcoh D}$ (up to prefactors),
as known before.
For $d_{x^2-y^2}$-wave hybridization, the two bands cross along the nodal lines,
but this crossing is at a finite energy away from the Fermi level.
Hence, the direct gap is finite and given by the renormalized $f$ level
position, $|\lambda_0|$ -- this translates into $\Delta_{\rm opt} \propto \Tcoh$.
Above this threshold energy, the optical conductivity follows $\sqrt{\omega-\Delta_{\rm opt}}$,
see App.~\ref{sec:OPLFK}.

As discussed above, a hard gap in $\sigma(\omega)$ will not survive beyond mean-field,
but we expect the qualitative result to remain valid.
We therefore conclude that a hybridization $\Vk$ with momentum-space nodes leads
to transfer of optical spectral weight from the energy scale $\sqrt{\Tcoh D}$ to the scale $\Tcoh$
(when compared to the case of constant hybridization).
For actual experiments this likely implies that no hybridization gap will be visible
in the optical conductivity, due to the finite width of the Drude peak.
Such a scenario is qualitatively consistent with the optical-conductivity data
obtained on CeMIn$_5$.\cite{burch}

\subsection{Thermal transport}
\label{sec:thermal}

Low-temperature d.c. transport quantities are in principle candidates to probe strong
anisotropies in momentum space.
As an example, let us consider the thermal conductivity
(which sometimes shows less sample dependence than the electrical conductivity).
The energy current operator in the mean-field approximation reads\cite{moreno1}
\begin{equation}
{\bf j}_T= \sum_{{\bf k} \sigma}
\left(
z_{1 {\bf k} }(\nabla_{\bf k} z_{1 {\bf k} })\gamma_{1 {\bf k}}^{\dagger}  \gamma_{1 {\bf k}}^{\phantom{\dagger}} +
(1 \leftrightarrow 2)
\right)
\end{equation}
where $\gamma_{1,2}^{\dagger}$ are the operators creating a quasi-particle in the $z_{1,2}$ band.
>From the Kubo formula, one derives the low-temperature thermal conductivity in relaxation-time approximation:
\begin{equation}
\kappa_{ij}= \frac 1 T
\sum_{{\bf k} \sigma} z_{2 {\bf k}}^2 (\nabla_{\bf k} z_{2 {\bf k}})_i (\nabla_{\bf k} z_{2 {\bf k}})_j   \left(-n_F'(z_{2 {\bf k}})\right)  \frac{1}{\Gamma (z_{2 {\bf k}})}
\end{equation}
where $\Gamma$ denotes the impurity-induced quasiparticle scattering rate,
and we have again assumed that only the band $\zb$ crosses the Fermi level.

As already discussed by Moreno and Coleman\cite{moreno1} in the context of gap-anisotropic Kondo insulators,
the thermal conductivity will be strongly anisotropic for 3d systems
where the hybridization has e.g. line nodes.
In contrast, in the 2d case of a $d_{x^2-y^2}$ hybridization, the conductivity tensor
does not have enough degrees of freedom to reflect the anisotropy, as the two principal
axes are equivalent here. (The sign of $\beta_{\bf k}$ does not enter.)
The same applies to hybridization function $V_{\bf k}$ with higher angular
momenta $l$.
Hence, for 2d anisotropic systems,
higher-order correlation functions need to be considered,
as e.g. probed by angle-dependent magnetoresistance;
this is beyond the scope of this paper.
[An exception is a $p$-wave hybridization (i.e. $l=1$)
which explicitly breaks the $C_4$ rotation symmetry down to $C_2$,
leading to an in-plane transport anisotropy.
Note that such a hybridization will be accompanied by a corresponding
lattice distortion, which will be reflected in the entire band structure.]

Finally, we note that, independent of possible transport anisotropies,
the Wiedemann-Franz law will always be obeyed (assuming elastic scattering
only):
The Lorenz number $L$, formed from the thermal conductivity $\kappa_{ii}$
and the electrical conductivity $\sigma_{ii}$ via $L=\kappa/(\sigma T)$,
will approach the constant $L_0 = (\pi^2/3) (k_B/e)^2$
in the low-temperature limit.
This is consistent with the fact that we are describing a Fermi liquid.
As a corollary, the recently observed violation of the Wiedemann-Franz law
in CeCoIn$_5$ at its field-induced critical point\cite{tanatar}
is likely related to inelastic scattering processes.


\section{Beyond mean-field theory}
\label{sec:beyond}

So far, we have discussed the low-temperature properties of ``unconventional''
heavy Fermi liquids using slave-boson mean-field theory.
In principle, corrections to mean-field theory can be systematically taken
into account, by considering fluctuations around the saddle point.
For the Kondo model, the correct implementation of the Hilbert space constraint,
together with phase fluctuations of the boson field, lead to
a theory where $\widetilde f$ and $b$ particles are minimally coupled to a
compact U(1) gauge field.
The Fermi-liquid phase corresponds to the Higgs/confining phase of the gauge
theory, it is stable w.r.t. fluctuation effects, their main effect being to endow the
$\widetilde f$-particle with a physical electric charge.\cite{read,millee}

To treat the full crossover from energies or temperatures above $\TK$ to those below $\Tcoh$,
different methods need to be employed.
Local correlations can be efficiently captured by dynamical mean-field
theory (DMFT).\cite{dmft-rmp}
If DMFT is formulated for the Anderson lattice model \eqref{eq:ALM}, correlation effects
arise from the {\em local} Hubbard interaction $U$, and consequently DMFT
can be used to treat an Anderson model with non-local hybridization $\Vk$ as well.
The DMFT self-consistency equation then reads:
\begin{equation}
\begin{split}
G_{\text{ALM,loc}}&(z)=\sum_{\bf k} \frac{1}{z-\epsilon_f-\Sigma_f(z)-\frac{V_{\bf k}^2}{z-\bar\epsilon_{\bf k}}}\\
&=\frac{1}{z-\epsilon_f-\widetilde \Delta(z)-\Sigma_f(z)}=G_{\text{SIAM}}(z)
\end{split}
\label{eq:dmft}
\end{equation}
Here, $\Sigma_f$ is the so-called interaction self-energy arising from $U$,
and $\widetilde\Delta$ denotes the effective hybridization function defined by
the second line of Eq.~\eqref{eq:dmft}.
While we shall not numerically solve the DMFT problem \eqref{eq:dmft} here,
we can briefly discuss a few properties.
Most importantly, the momentum dependence of the arising effective hybridization
is dictated by the bare $\Vk$.
This implies that all qualitative statements in Sec.~\ref{sec:FL} remain valid,
in particular all local moments will be fully screened at low $T$.
(Technically, the DMFT reduces the lattice model \eqref{eq:ALM} to an effective
single-impurity model, with a bath having a finite density of states at the Fermi
level -- this implies a fully developed Kondo effect as $T\to 0$.)

Cluster extensions of DMFT allow to handle momentum-dependent self-energies.
Then, in principle the momentum dependence of the effective hybridization will
differ from that of the bare $\Vk$. However, we do not expect qualitative changes
of the low-temperature physics described above.

Let us note one caveat:
While calculating thermodynamics and single-particle properties within DMFT
for the Anderson model \eqref{eq:ALM} is straightforward, electric transport
is not.
The reason is that the current operator inevitably involves contributions from
the non-local hybridization, see discussion in Sec.~\ref{sec:opcon}.
As a result, vertex corrections do not vanish in the DMFT limit,
in contrast to standard DMFT applications.\cite{dmft-rmp}


\section{Temperature-dependent resistivity}
\label{sec:rho}

In this section, we touch upon electronic properties at elevated temperatures.
In particular, we want to focus on the electrical resistivity $\rho(T)$ of
heavy fermions with unconventional hybridization for $T>\TK$.

In the conventional heavy-fermion picture, the electrical resistivity $\rho(T)$
at high temperatures, $T\gg\TK$, is small (ignoring phonons here),
and $\rho(T)$ rises upon lowering the temperature due to increasing magnetic
scattering.
At a scale which is often identified with the lattice coherence temperature $\Tcoh$,
$\rho(T)$ reaches a maximum and then drops upon further cooling,
behaving as $\rho(T) = \rho_0 + A T^2$ at low $T$.
At elevated temperatures, $T>\TK$, the scattering can be accessed using perturbation theory
in the Kondo coupling, i.e., the physical picture is that of $c$ electrons
(with a small Fermi surface) scattering inelastically off the $f$ moments.

Bare perturbation theory gives a single-particle scattering rate
\begin{equation}
\tau_{\bf k}^{-1} \propto J_0^2 \beta_{\bf k}^2 \left(1+ \frac{J_0}{D} \ln \frac D T\right).
\label{rate}
\end{equation}
A few remarks are in order:
(i) In the paramagnetic phase of a Kondo lattice,
all contributions to the conduction-electron self-energy up to order $J_0^3$
arise from single-impurity scattering.
(ii) The prefactor $\beta_{\bf k}^2$ comes from the two external lines of the
self-energy diagrams
whereas the internal momentum summations average out all other form factors --
this is also true for higher-order diagrams.
Assuming that scattering arises from the local moments {\em only},
the simplest approximation for the conductivity $\sigma\propto 1/\rho$ yields
\begin{equation}
\sigma_{ij}(T) \propto \int_{\rm FS} \frac{d^{d-1}k}{(2\pi)^d} \, v_i({\bf k}) v_j({\bf k}) \tau_{\bf k}
\label{conduc}
\end{equation}
where $v_i({\bf k}) = d \epsilon_{\bf k}/ d k_i$ is the quasiparticle velocity.
The result (\ref{rate},\ref{conduc}) is interesting, as it shows that for form factors $\beta_{\bf k}$
with nodes, the Kondo scattering is {\em insufficient} to render the conductivity finite,
because $\tau_{\bf k}$ diverges at least like $(k-k_n)^{-2}$ near the node at $k_n$.
The physical origin is that conduction electrons with momenta at the hybridization nodes
are not scattered at all,
and this short-circuits all other processes, leading to {\em infinite} conductivity.
To obtain a finite conductivity, additional scattering needs to be considered,
namely electron--electron scattering among the conduction electrons, electron--lattice
scattering, or scattering off static impurities.
The resulting interplay of scattering mechanisms can be complex and can even modify the
basic temperature dependence of $\rho(T)$,
but we shall not analyze it here in detail.

The physical conclusion is that the electrical current, at least in the temperature
regime $T>\TK$, is primarily carried by conduction electrons with weak hybridization
to the $f$ moments, hence ``nodal'' quasiparticles dominate the electric transport.
Recall that, in contrast, the low-temperature thermodynamics is dominated by ``antinodal''
quasiparticles.


\section{Competition between Kondo screening and ordering}
\label{sec:SC}

As already discussed by Doniach,\cite{doniach} the phase diagram of heavy-fermion metals
is determined by the competition between Kondo effect and inter-moment exchange
(either of direct or RKKY type).
Inter-moment exchange can drive magnetic ordering, but may also lead to
non-trivial metallic spin-liquid states and to magnetically mediated
superconductivity.
The competition with Kondo screening may be simply understood by stating
that the local $f$ moments can either form Kondo singlets with the conduction
electrons, or they can order in a symmetry-breaking fashion or else pair into
inter-moment singlets.

Thinking about these competing tendencies in momentum space,
it is conceivable that,
in a situation with momentum-space differentiation of electronic band properties,
certain ordering phenomena are favored or disfavored by a given form of the
hybridization.
This idea will be illustrated in this section, using
magnetically mediated superconductivity as an example,
where one can expect an intricate interplay between hybridization and
pairing symmetries.
Concrete calculations will be performed for a Kondo-Heisenberg model in
a mean-field approach:
As in Ref.~\onlinecite{svs}, a magnetic interaction between the $f$ moments
can be decoupled in the particle--particle channel, leading to
pairing of spinons, which, if coexisting with Kondo screening, leads
to BCS-type superconductivity.

\subsection{Kondo-Heisenberg model}

The Anderson and Kondo lattice models (\ref{eq:ALM},\ref{eq:KLM}) contain
the competition of Kondo and RKKY interactions.
However, in the slave-boson approach the effect of the RKKY interaction is lost.
For mean-field calculations it is thus convenient to introduce an explicit
inter-moment exchange interaction of Heisenberg type:
\begin{equation}
\mathcal{H}_{\rm H}=\sum_{i j}\frac{J_{H,ij}}{2}{\bf S}_i  \cdot{\bf S}_j \,.
\end{equation}

The physics of the model $\mathcal{H}_{\rm KLM}+\mathcal{H}_H$, commonly referred to as
Kondo-Heisenberg model, has been extensively discussed in the literature.
We shall give a comprehensive discussion of all phases and phase diagrams,
but instead concentrate on the possible emergence of superconductivity due to
$f$ electron pairing.
A general framework has been laid out in Ref.~\onlinecite{svs,svs2}, which considered
a scenario where dominant RKKY interaction does not lead to antiferromagnetism,
but instead to a metallic spin liquid state.
This state, arising e.g. from geometric frustration of the inter-moment exchange,
has been dubbed ``fractionalized Fermi liquid'' (FL$^\ast$), as it
features light conduction electrons, forming a Fermi liquid, which coexist
with a fractionalized spin liquid formed out of the $f$ electrons.
Then, in the generalized Doniach phase diagram, the heavy Fermi liquid (FL)
is separated from FL$^\ast$ by a quantum critical point where Kondo screening
breaks down, but no local symmetries are broken.\cite{svs,svs2}
(This quantum critical point has been discussed in relation to unconventional
quantum criticality in materials like CeCu$_{6-x}$Au$_x$ and YbRh$_2$Si$_2$.)

A specific realization of FL$^\ast$ is a state with paired spinons and
an emergent Z$_2$ gauge structure. As detailed in Ref.~\onlinecite{svs},
one can expect magnetically mediated superconductivity close to the quantum critical
point between FL and a Z$_2$ FL$^\ast$.
All resulting low-temperature phases can be conveniently described in a mean-field approach,
where the standard slave-boson description of the Kondo effect is combined
with a Sp($2N$) mean-field treatment of the Heisenberg exchange.\cite{sr}
Below, we shall extend this mean-field theory
to the case of momentum-dependent hybridization.

\subsection{Mean-field theory and magnetically mediated superconductivity}
\label{sec:mf_SC}

A mean-field theory for the Kondo-Heisenberg model,
$\mathcal{H}_{\rm KLM}+\mathcal{H}_{\rm H}$,
involves a decoupling of the Kondo interaction as in Sec.~\ref{sec:MF}
and of the inter-moment Heisenberg exchange $\mathcal{H}_H$.
Using the pseudofermion representation of the local moments as above,
non-local spinon pairing is described by a mean field of the form
$\Delta_{ij}=-\langle \widetilde
f_{i\uparrow}^{\phantom{\dagger}}\widetilde f_{j\downarrow}^{\phantom{\dagger}}-
\widetilde f_{i\downarrow}^{\phantom{\dagger}} \widetilde
f_{j\uparrow}^{\phantom{\dagger}}\rangle$.
Then, the Heisenberg interaction can be written at the mean-field level as\cite{sunfoot}
\begin{equation}
\mathcal{H}_{H,\text{MF}}= -\sum_{ij}\frac{J_{{\rm H},ij}}{4}\left[ \left( 2\Delta_{ij}
\widetilde f_{i\uparrow}^{\dagger}\widetilde  f_{j
\downarrow}^{\dagger}+h.c.\right)-|\Delta_{ij}|^2\right]\,.
\label{mfspn}
\end{equation}
For time-reversal invariant states, the bond field $\Delta_{ij}=\Delta_{ji}$ can be chosen to be real.
Importantly, the mean-field Hamiltonian \eqref{mfspn} is the exact solution of the
Heisenberg model in the symplectic Sp($2N$) large-$N$ limit, with a fully antisymmetric
representation of the local moments.\cite{sr}
(This large-$N$ limit uniquely selects the particle--particle decoupling of the Heisenberg
interaction.\cite{sunfoot})
Physicswise, non-zero $\Delta_{ij}$ creates a paramagnetic phase out of the $f$ moments;
in particular, uniform $\Delta_{ij}$ describes a gapped Z$_2$ spin liquid,
but also states with broken translational symmetry can occur which
can be classified as valence-bond solids.\cite{transnote}
A consistent Sp($2N$) mean-field treatment of the full model $\mathcal{H}_{\rm KLM}+\mathcal{H}_{\rm H}$
is obtained by also decoupling the Kondo interaction in the particle--particle channel.
However, one can show that for the Kondo part both particle--hole and
particle--particle decoupling schemes are equivalent regarding physical observables,
provided that $n_f=1$ and time-reversal symmetry is present.

The full mean-field theory is now given by
$\mathcal{H}_{\text{KLM,MF}}+\mathcal{H}_{\text{H,MF}}$, with the two
``order parameters'' $b$ and $\Delta$.
Restricting ourselves to states without translational symmetry breaking,
the following mean-field phases occur:
At high temperatures, a trivial decoupled phase with $b=\Delta=0$ is realized.
If the Heisenberg exchange $J_H$ dominates over the Kondo coupling $J_0$,
then $\Delta$ will be finite and $b$ zero at low $T$, resulting in decoupled $c$ and $f$ electron
subsystems: This is the fractionalized Fermi liquid FL$^*$ described above.
On the other hand, non-zero $b$ and vanishing $\Delta$ describe a conventional
heavy Fermi liquid FL (which was the subject of Sec.~\ref{sec:MF}).
Finally, if both $\Delta$ and $b$ are non-zero, Kondo screening coexists with
spinon pairing, which leads to a true superconducting state (SC), with
pairing mediated by the magnetic coupling among the $f$ moments.
At sufficiently low $T$, the FL phase is always unstable towards superconductivity
in the presence of a non-zero $J_H$.
(Note that the FL$^\ast$ phase is not a superconductor, as the $\widetilde f$
particles do not carry a physical charge in the absence of Kondo
screening.)
Fluctuation corrections to mean-field theory will smear out the finite-temperature
transitions of the FL and FL$^\ast$ phases (the latter only in $d\!=\!2$ dimensions),
whereas the superconducting transition remains a true phase transition.\cite{svs}

At this point, a more detailed discussion of the spatial structure of the Heisenberg
interaction, described by $J_{{\rm H},ij}$, and of the resulting pairing is needed.
For nearest-neighbor exchange on the square lattice of $f$ moments, each unit
cell contains two bond variables, $\Delta_{ij}$.
A numerical solution shows that two types of saddle points exist
(provided that translational and time-reversal invariance are imposed\cite{transnote}),
namely a uniform (or ``extended $s$-wave'') solution with $\Delta_{ij}=\Delta$
on all links, and a ``$d_{x^2-y^2}$-wave'' solution with $\Delta_{ij}=\pm \Delta$ on
horizontal and vertical links, respectively.
We will also consider the case where the $J_{{\rm H},ij}$ only act on next-neighbor
diagonal bonds, which together with an alternating structure of the $\Delta_{ij}$
leads to a ``$d_{xy}$-wave'' mean-field solution of the Heisenberg part.
All cases can be written in momentum space as
\begin{equation}
\mathcal{H}_{\text{H,MF}}=\sum_{{\bf k}} \widetilde W_{{\bf k}} \left( \widetilde f_{{\bf k}
\uparrow}^\dagger \widetilde  f_{-{\bf k}\downarrow}^\dagger+h.c.\right) +J_H\mathcal{N}\Delta^2
\label{eq:H_H}
\end{equation}
with the abbreviation $\widetilde W_{{\bf k}}=-J_H\Delta \alpha_{{\bf k}}$,
and $\alpha_{\bf k}$ contains the ``form factor'' of the spinon pairing:
\begin{equation}
\alpha_{\bf k} = \left\{
\begin{array}{ll}
 \cos k_x+\cos k_y & \text{extended $s$-wave}\\
 \cos k_x-\cos k_y &\text{$d_{x^2-y^2}$-wave}\\
2 \sin k_x \sin k_y & \text{$d_{xy}$-wave}\\
\end{array}
\right. .\label{eq:symmSC}
\end{equation}

The self-consistency equation which supplement the Hamiltonian
$\mathcal{H}_{\text{KLM,MF}}+\mathcal{H}_{\text{H,MF}}$, are given by
Eqs.~(\ref{eq:MF_KLM}) and
\begin{equation}
\sum_{{\bf k}} \alpha_{{\bf k}} \langle \widetilde f_{{\bf k} \uparrow}^\dagger \widetilde f_{-{\bf
k}\downarrow}^\dagger+h.c. \rangle=2 \mathcal{N} \Delta \,.
\label{eq:MF4}
\end{equation}
All expectation values can again be expressed in terms of Green functions, see App.~\ref{sec:GFH}.
Diagonalizing the mean-field Hamiltonian, we obtain the quasiparticle energies:
\begin{multline}
\bar z_{1,2,3,4\bf k} = \pm \frac{1}{\sqrt{2}} \left( \lambda_0^2+\bar \epsilon_{\bf
k}^2+2 b^2 \beta_{\bf k}^2+\widetilde W_{{\bf k}}^2 \phantom{\left(+\left(
\lambda_0^2+\bar \epsilon_{\bf k}^2+2 b^2 \beta_{\bf k}^2+\widetilde W_{\bf
k}^2\right)^2\right)^{1/2}}\right.\\ \phantom{1111111111} \left. \pm\left(-4
\left(\lambda_0 \bar \epsilon_{\bf k}+b^2 \beta_{\bf k}^2\right)^2-4\bar \epsilon_{\bf
k}^2 \widetilde W_{\bf k}^2 \phantom{\left( \lambda_0^2+\bar \epsilon_{\bf k}^2+2 b^2
\beta_{\bf k}^2+\widetilde W_{\bf k}^2\right)^2} \right.\right.\\ \left.\left.+\left(
\lambda_0^2+\bar \epsilon_{\bf k}^2+2 b^2 \beta_{\bf k}^2+\widetilde W_{\bf
k}^2\right)^2\right)^{1/2}\right)^{1/2}.
\end{multline}

Let us point out an interesting feature of the superconducting state SC:
Due to the momentum-dependent hybridization, the internal structure of the
Cooper pairs is highly non-trivial.
In particular, the anomalous expectation value of the physical electrons
is given by
$\langle c_{{\bf k}\uparrow}^{\dagger} c_{-{\bf k}\downarrow}^\dagger\rangle
\propto \Delta \beta_{\bf k}^2 \alpha_{\bf k} M_{\bf k}$ where
$M_{\bf k}$ is a smooth function respecting the lattice symmetries (see App.~\ref{sec:GFH}).
This transforms under the same representation of the space group as
the $\langle \widetilde f_{{\bf k}\uparrow}^{\dagger} \widetilde f_{-{\bf k}\downarrow}^\dagger\rangle
\propto \Delta\alpha_{\bf k}$ of the spinons, but has additional zeros from $\beta_{\bf k}^2$.
(A somewhat similar situation appears in the composite pairing picture
of Ref.~\onlinecite{productsc}, however, there arising from two-channel Kondo physics.)
The additional zeros will have a strong influence on thermodynamics, e.g.,
the power-law in the low-temperature specific heat will be modified.

\subsection{Qualitative discussion: $\Tc \ll \TK$ vs. $\Tc \sim \TK$}

Pairing in a Kondo-lattice system can occur in qualitatively different
regimes, depending on the relation between the superconducting $T_c$ and
the characteristic Kondo scale.
If $\Tc \ll \TK$, then the proper picture is that of BCS-like pairing
out of a well-formed heavy Fermi liquid.
In contrast, $\Tc \sim \TK$ implies a strong competition of Kondo screening
and Cooper pairing, and superconductivity emerges out of an incoherent
non-Fermi liquid regime.
(The formal situation $\Tc \gg \TK$ leads to the non-superconducting FL$^\ast$ phase.)

Let us quickly discuss the two regimes in the framework of the mean-field
theory, keeping in mind that inelastic processes at energies of order $\TK$
will not be captured.
The mean-field equation (\ref{eq:MF4}) reduces to an equation for $\Tc$
if we set $\Delta$ to zero:
\begin{equation}
\frac{2}{J_{\rm H}}=\frac{1}{\mathcal N}\sum_{\bf k} \alpha^2_{\bf k} \left(
\frac{\za^2-\bar \epsilon_{\bf k}^2}{\za^2-\zb^2} \frac{1}{\za} \tanh \frac{\za}{2 T_c} +
(1 \leftrightarrow 2)
\right)
\label{eq:Tc}
\end{equation}
The factor $\alpha^2_{\bf k}$ originates from the fact that the momentum dependence
of both the gap and the pairing interaction $\widetilde W_{\bf k}$ are equal by
construction.

In the regime $\Tc \ll \TK$ only the quasiparticle band crossing the
Fermi level (which we again assume to be $\zb$) contributes to pairing.
We replace the $\bf k$ summation by an integral over isoenergetic lines $\int d\omega
\int_{\omega=\zb} d A_{\bf k}/|\nabla \zb|$. In analogy to the standard BCS case we
approximate $\tanh (\omega/2 T_c)=1$ for $\omega \ge 2 T_c$ and 0 elsewhere. The factor
of $\alpha_{\bf k}^2$ will be taken at its value at the Fermi surface. Furthermore we
will set $\kappa \equiv (\bar \epsilon_{\bf k}^2-\zb^2)/(\za^2-\zb^2) \approx
\text{const}$, assuming that it is weakly varying along the Fermi surface.
Neglecting the dependencies of the velocity and $\alpha_{\bf k}$ perpendicular
to the Fermi surface we obtain
\begin{equation}
\frac{2}{J_H}=\int_{2 T_c}^{\Lambda} \frac{d\omega}{\omega}
\int_{\text{FS}} \frac{d A_{\bf k}}{2\pi}\, \frac{\kappa \alpha_{\bf k}^2}{|\nabla_{{\bf k}} \zb|}
\end{equation}
with $\Lambda$ of order the bandwidth.
This gives a rough $T_c$ estimate of
\begin{equation}
T_c=\frac{\Lambda}{2}\exp\left(-\frac{4 \pi}{J_H \kappa\int d A_{\bf k} (\alpha_{\bf k}^2 / |\nabla \zb|}\right)
\end{equation}
This equation shows the direct interplay of the form factors:
$T_c$ is enhanced if the pairing is strong (large $\alpha_{\bf k}$)
in regions where quasiparticles are heavy and hence have large $f$ character.
In other words, ``antinodal'' regions along the Fermi surface with large hybridization
are more susceptible to pairing.

In contrast, in the regime of $T_c\sim \TK$ both bands contribute to pairing,
and one can expect large contributions to the integral in Eq.~\eqref{eq:Tc}
from the essentially flat parts of both bands $\zab$, which are present
in particular close to the nodal lines of the hybridization $\Vk$.
This simple argument illustrates the competition
between Kondo effect and pairing in the regime $T_c\sim \TK$:
``Nodal'' momentum-space regions with less hybridization are more susceptible to pairing --
this is opposite to the statement made above for $\Tc \ll \TK$!
One should, however, keep in mind that the mean-field theory has limited relevance
for the true physics at energies or temperatures of order $\TK$,
the key point being that electrons in the ``antinodal'' regions are
rather incoherent.
The emerging problem of the pairing of incoherent fermions is of fundamental
relevance and heavily debated for instance in the field of high-temperature superconductors,
but rather little solid knowledge exists about this highly interesting strong-coupling
phenomenon.

\subsection{Extended phase diagrams}

\begin{figure}[tb]
\resizebox{0.49 \textwidth}{!}{
\includegraphics{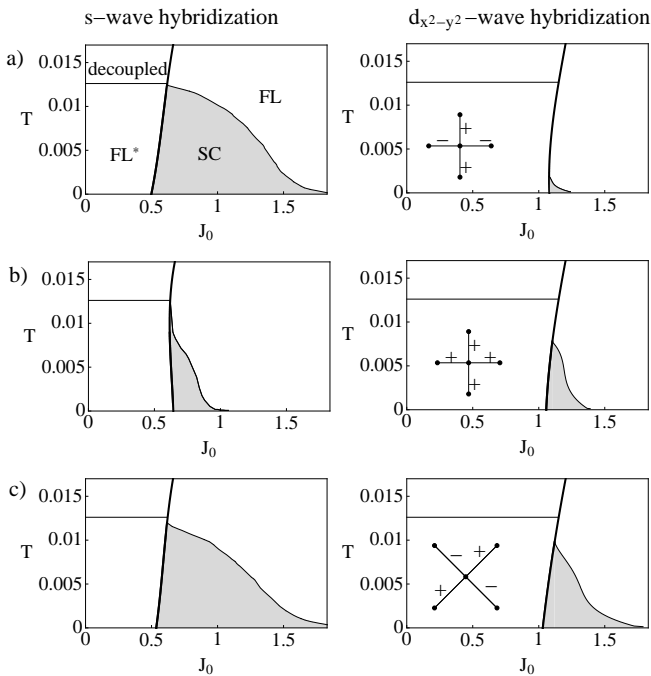}
}
\caption{
Phase diagrams in dependence of temperature $T$ and Kondo coupling $J_0$ at $n_c=0.3$
for an extended $s$- and a $d_{x^2-y^2}$-wave hybridization,
combined with different types of a superconducting symmetry:
a) $d_{x^2-y^2}$-wave,
b) extended $s$-wave, and
c) $d_{xy}$-wave.
The left-panel insets show the real-space structure of spinon pairing fields $\Delta_{ij}$,
leading to the certain type of pairing symmetry.
Thick (thin) lines refer to first (second) order phase transitions,
for further details see text.
}
\label{fig:PD}
\end{figure}

The discussion of the last section suggests that a certain hybridization symmetry can
favor or disfavor a certain pairing symmetry.
To follow up on this idea, we have determined mean-field phase diagrams from a fully
self-consistent numerical solution of Eqs.~\eqref{eq:MF_KLM} and \eqref{eq:MF4}.
While our results in principle support the above statement, they also show
that microscopic details of band structure, band filling, and pairing interaction
are important in determining $T_c$ (which may render simplistic arguments invalid).

\begin{figure}[b!]
\resizebox{0.5 \textwidth}{!}{
\includegraphics{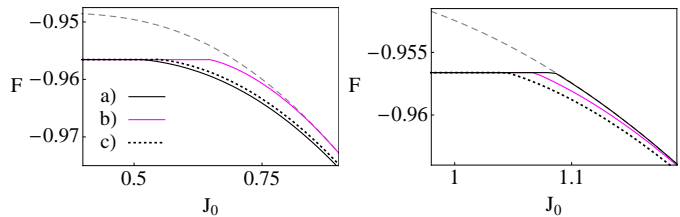}
}
\caption{
(Color online)
Free energy plotted against the Kondo coupling $J_0$ at $n_c=0.3$ and
$T=10^{-5}$ for extended $s$-wave hybridization (left) and $d_{x^2-y^2}$-wave
hybridization (right) and different superconducting symmetries:
a) $d_{x^2-y^2}$-wave,
b) extended $s$-wave, and
c) $d_{xy}$-wave
(compare to the corresponding phase diagrams shown in Fig. \ref{fig:PD})
The thin dashed line shows the normal-state solution with $\Delta=0$.
}
\label{fig:F}
\end{figure}

Sample phase diagrams as function of temperature $T$ and Kondo coupling $J_0$,
keeping $J_H$, $t$, and $n_c$ fixed, are shown in Fig.~\ref{fig:PD},
for hybridizations of extended $s$- and $d_{x^2-y^2}$-type and various pairing symmetries.
The overall structure of the phase diagram was discussed above in Sec.~\ref{sec:mf_SC}
and is identical to that described in Ref.~\onlinecite{svs}.

The different pairing symmetries in Fig.~\ref{fig:PD} have to be understood as follows:
For a given Heisenberg interaction, saddle points with different
spinon pairing symmetry occur, and to plot the phase diagrams we have restricted our
attention by hand to one of the saddle points.
The correct superconducting phase is obtained by comparing the free energies,
given by the mean-field expression
\begin{equation}
\begin{split}
F=-\frac{1}{\beta} &\sum_{{\bf k},i}\ln \left( 1+e^{-\beta \bar z_{i,\bf k}}\right)+\sum_{\bf k}(\epsilon_{\bf k}-\lambda_0)+\\
&+\mathcal{N} \mu n_c+\mathcal{N} \lambda_0+\mathcal{N} J_H \Delta^2+\mathcal{N}
\frac{b^2}{J_0}\,.
\end{split}
\end{equation}
Plots of the free energies at low $T$ are presented in Fig. \ref{fig:F}.

The rough conclusion for the particular dispersion and band filling used here
is that extended $s$-wave hybridization favors the $d_{x^2-y^2}$-wave superconducting
symmetry, while $d_{x^2-y^2}$-wave hybridization favors $d_{xy}$-wave superconductivity.
Given the structure of the Fermi surfaces in the FL phase, this is not unexpected:
If both $\alpha_{\bf k}$ and $\beta_{\bf k}$ have extended $s$ structure,
the pairing interaction is rather small near the Fermi surface (Fig.~\ref{fig:FF}b),
and consequently the superconductivity is weak in the left panel of Fig.~\ref{fig:PD}b,
whereas both $d$-wave pairing states perform nicely in energy.
For $\beta_{\bf k}$ of $d_{x^2-y^2}$ form, the Fermi surface (Fig.~\ref{fig:FF}c)
is mainly located close to the momentum-space diagonals, favoring
$d_{xy}$ pairing (right panel of Fig.~\ref{fig:PD}c).


\section{Conclusions}

In this paper, we have explored the consequences of a strongly
momentum-dependent hybridization between conduction and local-moment electrons
in heavy-fermion metals:
In the Fermi-liquid regime, the quasiparticle properties become strongly anisotropic
along the Fermi surface: ``nodal'' quasiparticles are light $c$ electrons,
whereas ``antinodal'' quasiparticles are heavy and have essentially $f$ character.
An interesting dichotomy arises: While the low-temperature thermodynamics
is dominated by heavy antinodal quasiparticles, the electrical conductivity
at elevated temperatures is carried by unhybridized nodal quasiparticles.
Experimentally important is the low-temperature optical conductivity $\sigma(\omega)$:
Due to the strongly momentum-dependent gap between the effective bands,
the hybridization gap in $\sigma(\omega)$ is essentially smeared out.

Further, we have advocated the idea that the momentum-space structure
of the hybridization is important in selecting ordering phenomena
which compete with Kondo screening near quantum criticality.
Here, two regimes need to be distinguished:
For energies or temperatures $T$ much smaller than the coherence temperature
$\Tcoh$, a weak-coupling quasiparticle picture is often appropriate,
and instabilities of the heavy Fermi liquid are determined by the
interaction among the (anisotropic!) quasiparticles.
In contrast, for $T\sim\Tcoh$ fascinating strong-coupling
phenomena can be expected, for example unconventional superconductivity
emerging from a non-Fermi liquid regime.
This physics will be dominated by inelastic processes, which again are strongly
anisotropic in momentum space.
A detailed study should be undertaken using cluster extensions of
dynamical mean-field theory, but is beyond the scope of this paper.

On the experimental side, CeNiSn and CeRhSb have been established to be
half-filled Kondo semimetals with a hybridization gap vanishing along
a certain crystallographic axis.\cite{ikeda,moreno1,kikoin}
The CeMIn$_5$ compounds are candidates for Kondo metals with strongly
anisotropic hybridization,\cite{burch} but other Ce or Yb materials where a clear-cut
hybridization gap in $\sigma(\omega)$ is absent may fall into this category as well.
We note that first-principles calculations based on density-functional theory
could, in principle, be able to
determine the hybridization symmetry, but strong interaction effects can render the
conclusions invalid. Recent x-ray absorption studies are promising in paving a way to an
experimental determination of the required microscopic information.\cite{tjeng}
To probe the anisotropic quasiparticle properties in the Kondo regime,
high-resolution angle-resolved photoemission is the ideal tool (with the restriction that
in can only be applied to quasi-2d systems).
As outlined in Sec.~\ref{sec:rho}, unusual behavior in the finite-temperature resistivity
may also be connected to nodes in the hybridization function.
Clearly, more detailed theoretical investigations of transport properties are needed.
Finally, we mention that the strong-coupling pairing regime $T_c \sim \TK$
is very likely realized in the fascinating superconductor PuCoGa$_5$.\cite{pucoga5}


\acknowledgments

We thank F. B. Anders, K. Burch, K. Haule, J. Paglione,
A. Rosch, T. Senthil, M. A. Tanatar, and V. Zlatic
for discussions.
This research was supported by the DFG through the SFB 608
and the Research Unit FG 960 ``Quantum Phase Transitions''.

~\\

{\it Notes added.}
While this paper was being completed, a related paper by Ghaemi {\em et al.}\cite{ghaemi2}
on angle-dependent quasiparticle weights appeared on the arXiv.
Their results for the low-temperature regime of anisotropic heavy Fermi liquids
are related to ours.

After submission of this paper, Shim {\em et al.}\cite{haule}
published a paper on first-principles calculations for CeIrIn$_5$,
which support the idea of a strongly momentum-dependent hybridization
function.


\appendix

\section{Mean-field theory}
\label{sec:GF}

In this appendix, we list the expressions of Green functions required for
the implementation of the mean-field theory.

\subsection{Green functions}
\label{sec:GF2}

The Kondo-lattice mean-field Hamiltonian can be rewritten in a matrix form:
\begin{equation}
\mathcal{H}_{\text{KLM}}= \sum_{{\bf k} \sigma} \Psi_{{\bf k}}^{\dagger} \underbrace{
\begin{pmatrix}
      \bar \epsilon_{\bf k} & b \beta_{\bf k} \\
      b \beta_{\bf k} & -\lambda_0 \\
    \end{pmatrix} }_{\hat H_{\bf k}}
\Psi_{{\bf k}}^{\phantom{\dagger}} +\text{const.}
\end{equation}
with $\Psi_{{\bf k}}=\left(c_{{\bf k} \sigma}^{\phantom{\dagger}}, \widetilde f_{{\bf k}
\sigma}^{\phantom{\dagger}}\right)^T$.
In the following, we shall denote retarded Green functions
\begin{equation}
\hat G_{AB}(z)=\int_0^{\infty} \text{d}t e^{i z t} \left(-i \theta(t) \langle [\hat
A(t),\hat B(0)]_+\rangle\right)
\end{equation}
as $\langle \langle \hat A;\hat B\rangle \rangle_z$.
Defining the matrix propagator
\begin{equation}
\hat G({\bf k},z) =\langle \langle\hat  \Psi_{\bf k}^{\dagger};\hat \Psi_{\bf
k}^{\phantom{\dagger}} \rangle \rangle =\left(z-\hat H_{\bf k}\right)^{-1}
\end{equation}
we obtain by explicit inversion
\begin{equation}
\begin{split}
\hat G({\bf k},z) &=
\begin{pmatrix}
 \langle \langle c_{{\bf k} \sigma}^{\dagger};c_{{\bf k} \sigma}^{\phantom{\dagger}}\rangle \rangle_z & \langle \langle c_{{\bf k} \sigma}^{\dagger}; \widetilde f_{{\bf k} \sigma}^{\phantom{\dagger}}\rangle \rangle_z\\
\langle \langle \widetilde f_{{\bf k} \sigma}^{\dagger}; c_{{\bf k}
\sigma}^{\phantom{\dagger}}\rangle \rangle_z& \langle \langle \widetilde f_{{\bf k}
\sigma}^{\dagger} ;\widetilde f_{{\bf k} \sigma}^{\phantom{\dagger}}\rangle \rangle_z\\
\end{pmatrix}\\
&=\frac{1}{(z-\za)(z-\zb)}
\begin{pmatrix}
      z+\lambda_0 & b \beta_{\bf k} \\
      b \beta_{\bf k} & z-\bar \epsilon_{\bf k} \\
\end{pmatrix}
\end{split}.
\end{equation}
The thermal expectation values required for the mean-field equation
are obtained by summing over Matsubara frequencies;
this can be done analytically, as the excitation energies, Eq.~(\ref{eq:excit}),
are known.

\subsection{Green functions in the presence of a Heisenberg term}
\label{sec:GFH}

The Hamiltonian containing the additional Heisenberg term,
Eq.~(\ref{eq:H_H}), has to be rewritten in a matrix form in analogy to
App.~\ref{sec:GF2}.
The inversion of $(z-\hat H_{\bf k})$ provides the needed Green functions.
We use the shorthand $h(z)=\prod_{i} (z-\bar z_i)$.
\begin{subequations}
\begin{equation}
\langle \langle \widetilde f_{{\bf k}\uparrow}^\dagger; \widetilde f_{{\bf k} \uparrow}^{\phantom{\dagger}}
\rangle \rangle_z=\frac{(z-\bar \epsilon_{\bf k})((z-\lambda_0) (z+\bar \epsilon_{\bf
k})-b^2 \gamma^2_k)}{h(z)}
\end{equation}
\begin{equation}
\langle \langle \widetilde f_{-{\bf k}\downarrow}^{\phantom{\dagger}}; \widetilde f_{-{\bf k}
\downarrow}^{\dagger} \rangle \rangle_z = \frac{(z+\bar \epsilon_{\bf k})((z+\lambda_0)
(z-\bar \epsilon_{\bf k})-b^2 \gamma^2_k)}{h(z)}
\end{equation}
\begin{equation}
\langle \langle \widetilde f_{{\bf k}\uparrow}^\dagger; \widetilde f_{-{\bf k} \downarrow}^{\dagger} \rangle
\rangle_z =
\langle \langle \widetilde f_{-{\bf k}\downarrow}^{\phantom{\dagger}}; \widetilde f_{{\bf k}
\uparrow}^{\phantom{\dagger}} \rangle \rangle_z =\frac{\widetilde W_{\bf k} (z^2-\bar
\epsilon_{\bf k}^2)}{h(z)}
\end{equation}
\begin{equation}
\langle \langle c_{{\bf k}\uparrow}^\dagger; c_{-{\bf k} \downarrow}^{\dagger} \rangle
\rangle_z =
\langle \langle c_{-{\bf k}\downarrow}^{\phantom{\dagger}}; c_{{\bf k}
\uparrow}^{\phantom{\dagger}} \rangle \rangle_z =-\frac{\widetilde W_{\bf k} b^2 \beta_{\bf k}^2}{h(z)}
\end{equation}
\begin{equation}
\begin{split}
\langle \langle c_{{\bf k}\uparrow}^\dagger; \widetilde f_{{\bf k} \uparrow}^{\phantom{\dagger}}
\rangle \rangle_z &=\langle \langle \widetilde f_{{\bf k}\uparrow}^\dagger; c_{{\bf k}
\uparrow}^{\phantom{\dagger}} \rangle \rangle_z\\ &=\frac{b \beta_{\bf k} ((z-\lambda_0)
(z+\bar \epsilon_{\bf k})-b^2\beta_{\bf k}^2)}{h(z)}
\end{split}
\end{equation}
\begin{equation}
\begin{split}
\langle \langle \widetilde f_{-{\bf k}\downarrow}^{\phantom{\dagger}}; c_{-{\bf k}
\downarrow}^{\dagger} \rangle \rangle_z &=\langle \langle c_{-{\bf
k}\downarrow}^{\phantom{\dagger}}; \widetilde f_{-{\bf k} \downarrow}^{\dagger} \rangle \rangle_z\\
&=\frac{-b \beta_{\bf k} ((z+\lambda_0) (z-\bar \epsilon_{\bf k})-b^2 \beta_{\bf
k}^2)}{h(z)}
\end{split}
\end{equation}
\begin{equation}
\langle \langle c_{{\bf k}\uparrow}^\dagger; c_{{\bf k} \uparrow}^{\phantom{\dagger}}
\rangle \rangle_z =\frac{-b^2 \beta_{\bf k}^2 (z+\lambda_0)+(z+\bar \epsilon_{\bf
k})(z^2-\widetilde W_{\bf k}^2-\lambda_0^2)}{h(z)}
\end{equation}
\begin{equation}
\langle \langle c_{-{\bf k}\downarrow}^{\phantom{\dagger}}; c_{-{\bf k}
\downarrow}^{\dagger} \rangle \rangle_z =\frac{-b^2 \beta_{\bf k}^2 (z-\lambda_0)+(z-\bar
\epsilon_{\bf k})(z^2-\widetilde W_{\bf k}^2-\lambda_0^2)}{h(z)}
\end{equation}
\end{subequations}

\section{Equivalence of Anderson and Kondo mean-field theories}
\label{sec:equiv}

Here we compare the two sets of mean-field equations
for the Anderson and Kondo lattice models, Eqs.~\eqref{eq:MF_ALM} and \eqref{eq:MF_KLM}.
These are expected to be equivalent, once the Kondo limit is
taken in the Anderson-model equations.
For $U\to\infty$, the Kondo coupling is $J_0 = V^2/|\epsilon_f|$.
Further, the Kondo limit implies $r\rightarrow 0$, for otherwise the
{\em effective} hybridization would diverge.
The average $\bar f$ occupation then becomes unity, and
the $\bar f$ and $\widetilde f$ operators are equivalent.
The physical valence fluctuations in the Anderson model are projected out by
$\epsilon_f \rightarrow -\infty$.
In this limit, the effective $\bar f$ level energy, $(\epsilon_f-\lambda)$, stays finite
(i.e. a fraction of the bandwidth) to ensure
$\sum_\sigma\langle\bar f_\sigma^\dagger \bar f_\sigma\rangle = 1$.
Therefore, $V^2/|\lambda|\rightarrow J_0$.

With this knowledge about the limiting behaviors,
the first of the Anderson-model mean-field equations (\ref{eq:MF_ALM}a) transforms like
\begin{equation}
\begin{split}
r V&= \frac{1}{2 \mathcal{N}} \langle \sum_{{\bf k}\sigma}\frac{V^2 \beta_{\bf
k}}{\lambda} \left(\bar f_{{\bf k} \sigma}^{\dagger} c_{{\bf k}
\sigma}^{\phantom{\dagger}}+c_{{\bf k} \sigma}^{\dagger} \bar f_{{\bf k}
\sigma}^{\phantom{\dagger}}\right)\rangle\\ &\rightarrow -\frac{J_0}{2\mathcal{N}}\langle
\sum_{{\bf k} \sigma}\beta_{\bf k} \left( \widetilde f_{{\bf k} \sigma}^{\dagger} c_{{\bf k}
\sigma}^{\phantom{\dagger}}+c_{{\bf k} \sigma}^{\dagger} \widetilde f_{{\bf k}
\sigma}^{\phantom{\dagger}}\right)\rangle=b
\end{split}
\end{equation}
We can see that the mean-field equations of both theories correspond to each other,
and $r V$ and $b$ are the effective band hybridizations in the two-band model.

\section{Optical conductivity}
\label{sec:OPLFK}

We briefly discuss the behavior of the inter-band optical conductivity close to the
threshold energy, for the case of $d$-wave hybridization.
In Eq.~\eqref{eq:sigma1}, the matrix elements are non-singular near the threshold, hence
we only need to analyze the behavior of
\begin{equation}
\delta\left(\omega-\underbrace{\sqrt{(\lambda_0+\bar
\epsilon_{\bf k})^2+4 b^2 \beta_{\bf k}^2}}_{\Delta E(k_x,k_y)}\right)
\end{equation}
A change of variables is a:
\begin{subequations}
\begin{equation}
\widetilde k_x=\frac{1}{\sqrt{2}} (k_x+k_y-2 k_0)
\end{equation}
\begin{equation}
\widetilde k_y=\frac{1}{\sqrt{2}} (-k_x+k_y)
\end{equation}
\end{subequations}
$k_0$ is determined by $\epsilon(k_0,k_0)=0$ For $\omega$ near $-\lambda_0$, $\Delta
E(\widetilde k_x,\widetilde k_y)$ can be approximated by a second-order Taylor expansion
in $\widetilde k_x$ and $\widetilde k_y$ around 0. The $\bf k$ integration is restricted
to the Fermi sea because of the factor $n_F(z_{2 {\bf k}})$. The Fermi surface can
approximated by also expanding around  $\widetilde k_x$, $\widetilde k_y=0$.
Power counting in the integral then shows, that the first
non-vanishing contribution to the optical conductivity for a $d$-wave hybridization
is $\propto \sqrt{\omega+\lambda_0}$.



\begin{thebibliography}{}
\bibitem{heatfoot}
In the mean-field approximation, the free energy also acquires a
temperature dependence from the $T$ dependence of the mean-field
parameters. However, at low $T$, the corrections are quadratic in
$T$, leading to specific-heat contributions $\propto T^2$,
being subleading compared to $C\propto T$ from the fermionic
quasiparticles.\cite{hewson,readnewns}
Fluctuations around the mean-field solution do not change
the low-temperature specific heat: Amplitude fluctuations are
gapped, and phase fluctuations (of $r$ or $b$) do not exist,
as there is no physical phase degree of freedom.
(Formally, those fluctuations are gapped by the Anderson-Higgs
mechanism of the compact U(1) gauge field.)
\bibitem{sunfoot}
A decoupling of the Heisenberg interaction in the particle--hole channel
is possible as well. Then, in the limit $J_H\gg \TK$
a U(1) FL$^\ast$ phase with gapless spinons (i.e. an RVB-like spin liquid)
is realized, and superconductivity does not occur.\cite{svs2}
As our focus here is on symmetry-broken phases competing with
Kondo screening, we restrict ourselves to the particle--particle
decoupling (which is selected by the Sp($2N$) large-$N$ limit).
\end{thebibliography}
\end{document}